\documentclass[%
 reprint,
 amsmath,amssymb,
 aps,
]{revtex4-2}
\usepackage[utf8]{inputenc}

\usepackage{graphicx}
\usepackage{dcolumn}
\usepackage{bm}
\usepackage{subcaption}
\usepackage{amsmath}
\usepackage{siunitx}
\usepackage[utf8]{inputenc}
\DeclareUnicodeCharacter{0304}{}
\usepackage{soul} 
\usepackage[normalem]{ulem}
\usepackage{xcolor}
\usepackage{lineno}

\usepackage{etoolbox}
\usepackage{etoolbox}
\AtBeginEnvironment{table}{\nolinenumbers}
\AtBeginEnvironment{figure}{\nolinenumbers}
\AtBeginEnvironment{algorithm}{\nolinenumbers}
\AtEndEnvironment{table}{\linenumbers}
\AtEndEnvironment{figure}{\linenumbers}
\AtEndEnvironment{algorithm}{\linenumbers}
\makeatletter
\AtBeginEnvironment{subfigure}{\nolinenumbers}
\AtEndEnvironment{subfigure}{\linenumbers}
\makeatother
\definecolor{modblue}{RGB}{255, 255, 204} 
\sethlcolor{modblue} 
\begin{document}

\preprint{APS/123-QED}

\title{LHAASO Constraints on New-Source Cosmic Rays:\\
 Abundance Scenarios and Injection Spectral Indices}

\author{Xu-Lin Dong}
\affiliation{School of Physics, Hebei Normal University, Shijiazhuang, 050024 Hebei, China}
\affiliation{State Key Laboratory of Particle Astrophysics, Institute of High Energy Physics, Chinese Academy of Sciences, Beijing 100049, China}

\author{Pei-Pei Zhang}
\affiliation{School of Intelligence and Information Engineering, Tangshan University, Tangshan, 063000 Hebei, China}
\affiliation{State Key Laboratory of Particle Astrophysics, Institute of High Energy Physics, Chinese Academy of Sciences, Beijing 100049, China}
\affiliation{The Key Laboratory of Cosmic Rays, Xizang University, Ministry of Education, Lhasa 850000, Tibet, China}

\author{Qi-Ling Chen}
\affiliation{College of Physics, Sichuan University, Chengdu 610064, Sichuan Province, China}
\affiliation{State Key Laboratory of Particle Astrophysics, Institute of High Energy Physics, Chinese Academy of Sciences, Beijing 100049, China}

\author{Shu-Wang Cui}\email{cuisw@hebtu.edu.cn}
\affiliation{School of Physics, Hebei Normal University, Shijiazhuang, 050024 Hebei, China}
\affiliation{TIANFU Cosmic Ray Research Center, Chengdu, 610101 Sichuan, China}

\author{Yi-Qing Guo}\email{guoyq@ihep.ac.cn}
\affiliation{State Key Laboratory of Particle Astrophysics, Institute of High Energy Physics, Chinese Academy of Sciences, Beijing 100049, China}
\affiliation{University of Chinese Academy of Sciences, Shijingshan District Beijing 100049, China}
\affiliation{TIANFU Cosmic Ray Research Center, Chengdu, 610101 Sichuan, China}

\begin{abstract}

Recent LHAASO observations indicate that microquasars and massive star clusters contribute to Galactic cosmic rays, and likely also to the pre-knee spectral hardening. Their markedly different astrophysical environments, compared with supernova remnants (SNRs), may imprint distinct cosmic-ray abundances. We investigate the spectra and composition of new-source cosmic rays under two abundance scenarios: Case-1, the SNR-like composition with enhanced heavy nuclei, and Case-2, the surface abundances of microquasar companion stars. We constrain injection spectral indices using LHAASO proton, helium, and all-particle spectra, and use their ⟨lnA⟩ data for model discrimination. Under the assumption of a unified heavy-element injection spectrum, Case-2 fits the ⟨lnA⟩ data significantly better than Case-1, but at the cost of large, physically unmotivated differences between heavy-element and He injection indices. This suggests the need to either relax the unified heavy-element injection-index assumption or explore alternative abundance patterns. We predict that if the new-source composition follows Case-1, the individual elemental spectral indices before the knee should remain nearly consistent; if it follows Case-2, the Mg, Si, and Fe spectra will be significantly softer than those of He, C, and O. Future element-resolved energy-spectrum measurements by HERD and LHAASO will provide a decisive test of this discriminating feature.

\end{abstract}

\maketitle

\section{\label{sec:level1}Introduction}
Supernova remnants (SNRs) have long been considered the primary origin of Galactic cosmic rays\cite{2020PhR...872....1B}.
However, in recent years, experiments such as LHAASO have successively confirmed that massive star clusters\citep{2024SciBu..69..449L} and microquasars\citep{2025arXiv250514447T,2025NSRev..12af496L} can also efficiently accelerate cosmic rays.
In particular, the precise measurement of the proton energy spectrum by LHAASO has revealed a new prominent spectral hardening feature before the knee\citep{2026PhRvL.136l1001C,2025arXiv250514447T}, suggesting that this component may originate from non-SNR sources such as microquasars.
This strongly suggests that Galactic cosmic rays may be jointly contributed by at least two classes of astrophysical sources with distinctly different physical properties.
This paper aims to study the elemental abundances of cosmic rays contributed by new cosmic-ray sources, using experimental data on energy spectra and composition.

The prevailing picture of cosmic-ray acceleration by supernova remnants is that a forward shock sweeps through the cold interstellar medium.
In this process, the shock ionizes and heats the interstellar medium, and particles are accelerated through diffusive shock acceleration and contribute a power-law spectrum with a spectral index of about -2\cite{1988REVIEWS, drury1983introduction, dorman2006cosmic}.
Meanwhile, the systematic deviation between cosmic-ray elemental abundances and ISM abundances can be explained by two mechanisms: the first ionization potential (FIP) effect—low-FIP elements\citep{1997ApJ...487..182M}, being more easily ionized in the cold medium, preferentially enter the acceleration process, leading to their relative enhancement; and the dust-preferential injection mechanism—refractory elements (Mg, Si, Fe)\citep{2025A&A...693A.145C}, existing in the form of solid dust grains with extremely large mass-to-charge ratios (A/Q), thereby achieving higher injection efficiencies.
Together, these two constitute the “cold-medium chemical imprint” of SNR-like sources.

In contrast, the physical picture of microquasars typically involves a black hole accreting matter from a companion star (an OB star) and producing jets.
There are currently three main viewpoints regarding the specific mechanism by which microquasars accelerate cosmic rays\citep{1998NewAR..42..579A,2020ApJ...889..146S,2024JHEAp..43...93K}: (1) shear acceleration in the accretion disk; (2) acceleration of the hot ionized material within the jet by the termination shock at the jet end; (3) acceleration of the external cold interstellar medium by the forward shock at the jet head.
For viewpoints (1) and (2), the accelerated material originates mainly from the companion star and is highly ionized, so the physical premises of the first ionization potential effect and the dust-preferential injection theory no longer hold.
Therefore, if processes (1) and (2) dominate cosmic-ray acceleration in microquasars, their abundances should be close to those of OB stars.
For viewpoint (3), since the accelerated material is the same as that in SNRs, its abundance should be close to that of SNRs.
The acceleration process in massive star clusters (dominated by OB stars) is usually attributed to termination shocks formed by the collision of intense stellar winds.
Since the accelerated material is mainly stellar wind that has been heated and fully ionized, the physical conditions for the first ionization potential effect and the dust-preferential injection theory are likewise absent, and the abundances contributed by these sources should also be close to the OB-star composition.

In summary, the cosmic-ray abundances contributed by new cosmic-ray sources can be described by two extreme scenarios: Case-1 (SNR mode), in which both low-FIP elements and refractory elements are enhanced, leading to relatively high heavy-element abundances; and Case-2 (OB-star surface abundance), in which there is no element-selective enhancement, leading to relatively low heavy-element abundances.
This fundamental difference is the key to identifying the types and acceleration mechanisms of new cosmic-ray sources.

Recently, LHAASO has successively published precise measurements of the proton, helium, and all-particle energy spectra, together with the average logarithmic mass ⟨lnA⟩, near the knee. This makes it possible to constrain, using experimental data, the elemental abundances and injection spectral indices of cosmic rays contributed by new cosmic-ray sources.

In this work, we adopt two assumptions for the elemental abundances of cosmic rays contributed by new sources, namely Case-1 and Case-2, and require all elements other than protons and helium to share a unified injection spectral index. We constrain the spectral indices using the LHAASO proton, helium, and all-particle spectra, and use ⟨lnA⟩ as the decisive observable for model discrimination. 
The results show that, under equally good fits to the energy spectra, Case-2 provides a significantly better fit to ⟨lnA⟩ than Case-1, but yields dramatic differences among the injection spectral indices of heavy elements, and He that lack physical regularity.
This suggests that, under the current constraints on spectral indices, both extreme abundance assumptions are difficult to perfectly explain the observational data. It may therefore be necessary to relax the requirement of uniform injection spectra across components, or to adopt new elemental abundances.

Truly determining the abundances of new sources requires reliance on precise measurements of the element-resolved energy spectra before the knee.
We expect that if the new-source contribution follows Case-1, the spectral shapes of all elements before the knee will be basically consistent; if it follows Case-2, the spectra of Mg, Si, and Fe will be significantly softer than those of He, C, O, and Ne.
We anticipate that future experiments such as HERD and LHAASO will perform precise measurements of this feature to verify our conjecture.

The structure of this paper is as follows. Section 2 describes the propagation model, elemental abundance choices, data selection, and fitting procedure; Section 3 presents the results; Section 4 provides the conclusions and expectations.
The appendix provides detailed comparisons of individual elements under different abundance assumptions, as well as sensitivity analyses based on different hadronic interaction models.

\section{MODEL AND METHODOLOGY}

In the following, a quantitative description of the CR propagation process is provided. The propagation of CRs can be represented mathematically by the following partial differential equation\cite{maurin2002galactic}
\begin{equation}
\begin{split}
\frac{\partial\Psi(\bm{r},p,t)}{\partial t} = Q(\vec{r},p,t)+\vec{\nabla}\cdot(D_{xx}\vec{\nabla}\Psi-\vec{V}_{c}\Psi)+ \\
\frac{\partial}{\partial p}[p^{2}D_{pp}\frac{\partial}{\partial p}\frac{\Psi}{p^{2}}]-\frac{\partial}{\partial p}[\dot{p}\Psi-\frac{p}{3}(\vec{\nabla}\cdot\vec{V}_{c})\Psi]-\frac{\psi}{\tau_{f}}-\frac{\psi}{\tau_{r}},
\end{split}
\end{equation}
where $\Psi(\vec{r},p,t)$ denotes the cosmic-ray density, which is a function of position vector ($\vec{r}$), particle momentum ($p$), and time ($t$). $Q(\vec{r},p,t)$ describes the distribution of sources, $D_{xx}$ is defined as the spatial diffusion coefficient, $\vec{V}_{c}$ represents the convection velocity, $D_{pp}$ is the momentum diffusion coefficient accounting for the reacceleration process, $\dot{p}$, $\tau_{f}$ and $\tau_{r}$ are the momentum loss rate, the fragmentation timescale, and the radioactive decay timescale, respectively. 

{
\begin{table}[]
\caption{\label{tab:table1}Propagation model parameters.}
\begin{ruledtabular}
\begin{tabular}{cccccccccccccccc}
$D_0$ &$\delta_0$&$\quad N_m$&$\quad\xi$&$\quad n$&$\quad v_a$\footnotemark[1]&$\quad z_0$&$r_0$\ \\
$[\mathrm{cm}^{2}\mathrm{s}^{-1}]$&&&&&$[\mathrm{km~s}^{-1}]$&$[\mathrm{kpc}]$&$[\mathrm{kpc}]$\\ \hline
 $4.8\times10^{28}$&0.65&0.2&0.082&4.0&6&5&20 \\
\end{tabular}
\end{ruledtabular}
\footnotetext[1]{$v_a$ is the Alfvén velocity.}
\end{table}

\begin{table*}[!htb]
\caption{\label{tab:table2}Spectral injection parameters\protect\footnotemark}
\begin{ruledtabular}
\begin{tabular}{lcccccccc}
& \multicolumn{2}{c}{Abundance} & \multicolumn{2}{c}{Type-A} & \multicolumn{2}{c}{Type-B (Case-1)} & \multicolumn{2}{c}{Type-B (Case-2)} \\
\cline{2-3} \cline{4-5} \cline{6-7} \cline{8-9}
& Case-1 & Case-2 & Normalization & $\gamma$ & Normalization & $\gamma$ & Normalization & $\gamma$ \\
& & & $[\mathrm{GeV}^{-1}\mathrm{m}^{-2}\mathrm{s}^{-1}\mathrm{sr}^{-1}]$ & & $[\mathrm{GeV}^{-1}\mathrm{m}^{-2}\mathrm{s}^{-1}\mathrm{sr}^{-1}]$ & & $[\mathrm{GeV}^{-1}\mathrm{m}^{-2}\mathrm{s}^{-1}\mathrm{sr}^{-1}]$ & \\
\hline
H  & $1.29\times10^{6}$ & $1.29\times10^{6}$ & $4.00\times10^{-2}$ & 2.41 & $3.70\times10^{-3}$ & 2.15 & $3.70\times10^{-3}$ & 2.15 \\
He & 64600              & 235000              & $2.00\times10^{-3}$ & 2.27 & $1.85\times10^{-4}$ & 2.09 & $6.74\times10^{-4}$ & 2.20 \\
C  & 2700               & 515                 & $8.37\times10^{-5}$ & 2.28 & $7.74\times10^{-6}$ & 2.14 & $1.48\times10^{-6}$ & 2.00 \\
O  & 3500               & 1415                & $1.09\times10^{-4}$ & 2.32 & $1.00\times10^{-5}$ & 2.14 & $4.06\times10^{-6}$ & 2.00 \\
Ne & 540                & 296                 & $1.67\times10^{-5}$ & 2.29 & $1.55\times10^{-6}$ & 2.14 & $8.49\times10^{-7}$ & 2.00 \\
Mg & 790                & 90                  & $2.45\times10^{-5}$ & 2.36 & $2.27\times10^{-6}$ & 2.14 & $2.58\times10^{-7}$ & 2.00 \\
Si & 750                & 78                  & $2.33\times10^{-5}$ & 2.36 & $2.15\times10^{-6}$ & 2.14 & $2.24\times10^{-7}$ & 2.00 \\
Fe & 720                & 70                  & $2.23\times10^{-5}$ & 2.37 & $2.07\times10^{-6}$ & 2.14 & $2.01\times10^{-7}$ & 2.00 \\
\end{tabular}
\end{ruledtabular}
\footnotetext{The cutoff magnetic rigidity of Type-A sources is $R_{\rm cut,A}=35$~TV, $\alpha_A=0.5$ ($R_{\rm cut,A,Fe}=20$~TV, $\alpha_{A,Fe}=1$); that of Type-B sources is $R_{\rm cut,B}=3.5$~PV, $\alpha_B=0.75$.}
\end{table*}


\subsection{Source Injection}
The source term $Q(\vec{r},p,t)$ describes where CRs are injected in the Galaxy and with what spectrum. In this work, we consider two source components: Type-A sources, and Type-B sources. Type-A sources represent the conventional Galactic CR population, with acceleration limits of several hundred TeV per nucleon, as expected for typical supernova remnants. Type-B sources describe a higher-energy Galactic population capable of accelerating particles up to several PeV per nucleon. Recent LHAASO observations suggest that sources such as microquasars may reach this energy range \citep{2025NSRev..12af496L,2025arXiv251216638T}. The Type-A and Type-B source populations are treated as steady-state continuous source distributions.

In the steady-state approximation, the source term of each Galactic source population can be factorized into a spatial distribution and an injection spectrum,
\begin{equation}
Q_i(\vec{r},p)=f_i(r,z)q(R),
\end{equation}
where $i=A,B$, $f_i(r,z)$ gives the Galactic spatial distribution of the corresponding source population, and $q(R)$ describes its rigidity-dependent injection spectrum. The spatial distribution of Type-A sources is approximated by that of supernova remnants \citep{1996A&AS..120C.437C},
\begin{equation}
f_{A}(r,z)=
\left(\frac{r}{r_\odot}\right)^{1.25}
\exp\left[-\frac{3.87(r-r_\odot)}{r_\odot}\right]
\exp\left(-\frac{|z|}{z_s}\right),
\end{equation}
where $r_{\odot}=\SI{8.5}{kpc}$ and $z_s=\SI{0.2}{kpc}$. The spatial distribution of Type-B sources is approximated by that of X-ray binaries \citep{2026ApJ...997..163Y},
\begin{equation}
f_{B}(r,z)=
\frac{
\exp \left(-a\frac{r}{R_{\odot}}\right)
+B\exp \left(-b\frac{r^2}{R_{\odot}^2}\right)
}{
\exp(-a)+B\exp(-b)
}
\exp\left(-\frac{|z|}{z_s}\right),
\end{equation}
where $a=2.67\pm0.21$, $b=39.33\pm3.88$, and $B=9.74\pm1.99$.

The elemental abundances adopted for Type-A and Type-B sources are as follows:
For Type-A sources (representing supernova remnants), the abundances are based on the default settings in the DRAGON\citep{evoli2008cosmic} code, with necessary adjustments made according to the experimental data from AMS-02 and DAMPE. This set of elemental abundances is referred to as Case-1. 

For Type-B sources (representing microquasars, etc.), we consider two assumptions regarding their elemental abundances:

Case-1: Identical to the abundances of Type-A sources.

Case 2: consistent with the elemental abundances of B-type stars\citep{Nieva_2012,2008ApJ...688L.103P} (the companion stars of microquasars).

Both abundance sets are normalized using the hydrogen normalization coefficient of $N 
_H=1.29\times10^6$, with the detailed values listed in Table 2.

The injection energy spectra of elements adopt a sub-exponential cutoff.

\begin{equation}
q(R)\propto R^{-\gamma } \cdot \exp \left [-\left (  \frac{R}{R_c} \right )^{\alpha}  \right ] 
\end{equation}
Here, $R_c$ is the cutoff rigidity, $\alpha$ is sub-exponential index and $\gamma$ are the injection spectral indices, respectively.
This functional form is adopted to properly match the proton and helium energy spectra measured by DAMPE and LHAASO, as a standard exponential cutoff fails to accurately describe the spectral shape in this energy range.

The detailed injection parameters are listed in Table~II.

\subsection{Spatial Diffusion}
Charged CR particles are scattered by turbulent magnetic fields during propagation, and their transport is generally described as a diffusion process. The diffusion coefficient $D_{xx}$ characterizes the efficiency of this spatial transport. In conventional CR propagation models, $D_{xx}$ is usually assumed to be spatially homogeneous and to depend only on particle rigidity. However, observations of extended emission around CR sources by HAWC \citep{abeysekara2017extended} and LHAASO \citep{aharonian2021extended} suggest that diffusion in source environments can be much slower than that in the average interstellar medium (ISM). This motivates the use of spatially dependent propagation (SDP) models, in which the diffusion coefficient varies with position and is correlated with the Galactic source distribution.

The SDP model adopted in this work follows the framework developed and applied in previous studies to describe Galactic CR spectral anomalies, large-scale anisotropy, and diffuse gamma-ray emission; see \citep{yao2024common,2024arXiv241209016D,PhysRevD.109.063027,2026JCAP...07..050Q,2020FrPhy..1524601Y,2019SCPMA..6249511Y,2017PhRvD..95h3007Y,2026PhRvD.113l3037Y} for more details. In this framework, the diffusion region is divided into inner and outer zones. The diffusion coefficient in the inner zone is anticorrelated with the source distribution, while that in the outer zone follows the conventional spatially independent form. Mathematically, the diffusion coefficient is parameterized as\cite{2016ApJ...819...54G,PhysRevD.97.063008}
\begin{equation}
D_{xx}(r,z,R)=D_{0}F(r,z)\beta^{\eta}
\left(\frac{R}{R_{0}}\right)^{\delta_0F(r,z)},
\end{equation}
where $D_0$ is the normalization at the reference rigidity $R_0=\SI{4}~$GV, $\beta=v/c$, and $\delta_0$ controls the rigidity dependence of diffusion. The factor $\beta^\eta$ accounts for low-energy corrections; in this work this low-energy correction is switched off, i.e., $\eta = 0$. The spatial modulation function $F(r,z)$ is defined as
\begin{equation}
F(r,z)=
\begin{cases}
g(r,z)+[1-g(r,z)]
\left(\frac{|z|}{\xi z_0}\right)^n,
& |z|\le \xi z_0, \\
1,
& |z|>\xi z_0 ,
\end{cases}
\end{equation}
with
\begin{equation}
g(r,z)=\frac{N_m}{1+f(r,z)} .
\end{equation}
Here, $N_m$ controls the normalization of the diffusion suppression, $\xi$ gives the fractional thickness of the inner halo, and $n$ controls the smoothness of the transition between the inner and outer zones. In this work, the spatial modulation of the diffusion coefficient is assumed to follow the large-scale Galactic source distribution, which we approximate using $f(r,z)=f_A(r,z)$.

We solve the transport equation using the numerical package DRAGON \citep{evoli2008cosmic}. The interstellar gas distribution, CO distribution, and secondary fluxes are calculated using the GALPROP reaction cross sections embedded in DRAGON \citep{evoli2008cosmic}. Solar modulation is included using the force-field approximation \citep{gleeson1968solar}. The key SDP parameters adopted in this work are summarized in Table~I. 

{
\nolinenumbers 

\begin{figure}
    \centering
    \includegraphics[width=1\linewidth]{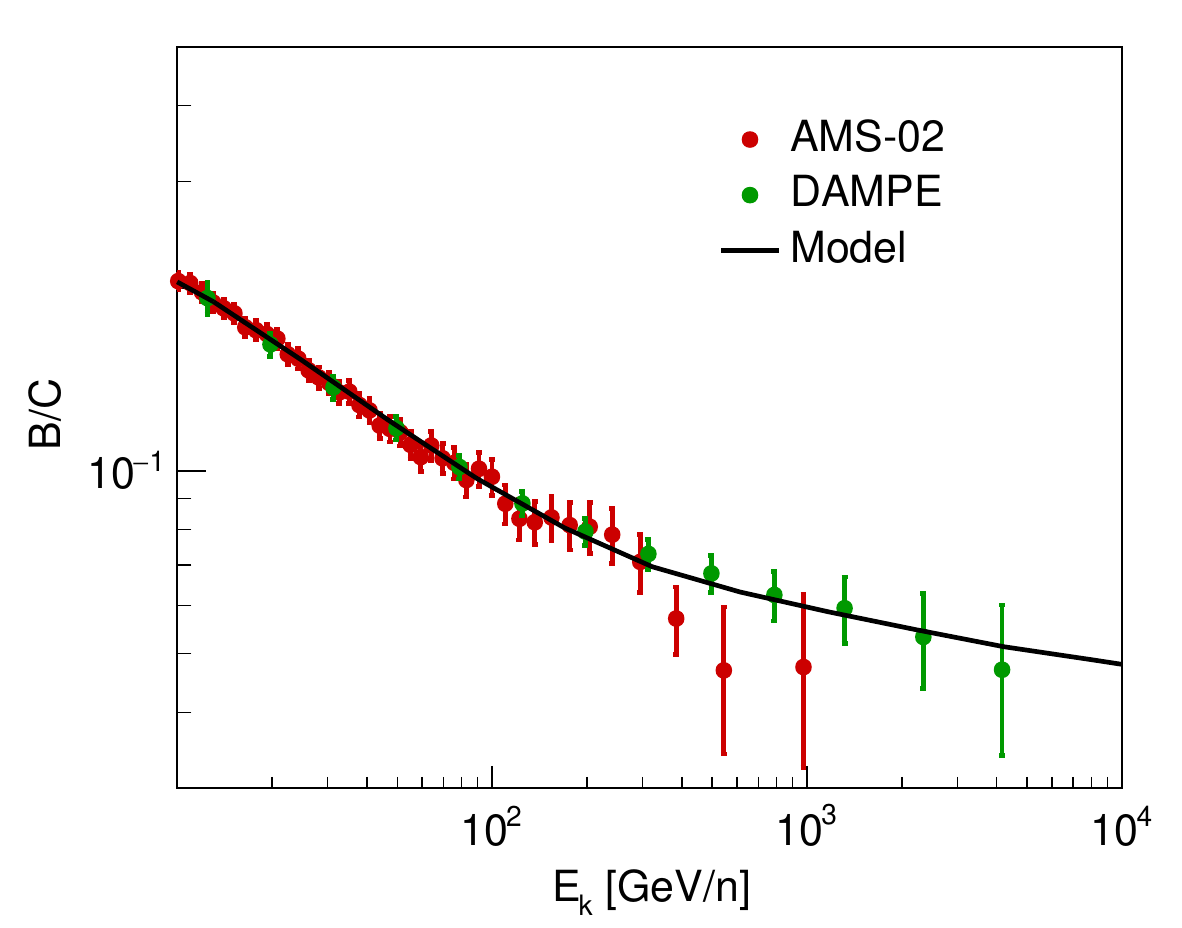}
    \caption{Comparison of the model-calculated B/C ratio with the AMS-02\cite{aguilar2021alpha} and DAMPE\cite{DAMPE:2022jgy} measurements. The solid black line denotes the model calculation; the red and green data points denote the AMS-02 and DAMPE measurements, respectively.
    }
    \label{fig:placeholder}
\end{figure}

\linenumbers} 

{
\nolinenumbers 
\begin{figure*}[]  
    \centering
    \begin{subfigure}[t]{0.46\textwidth}  
        \centering
        \includegraphics[
            width=\linewidth,
            height=6.6cm,                  
            keepaspectratio,             
            bb=0 0 500 400,              
            draft=false                  
        ]{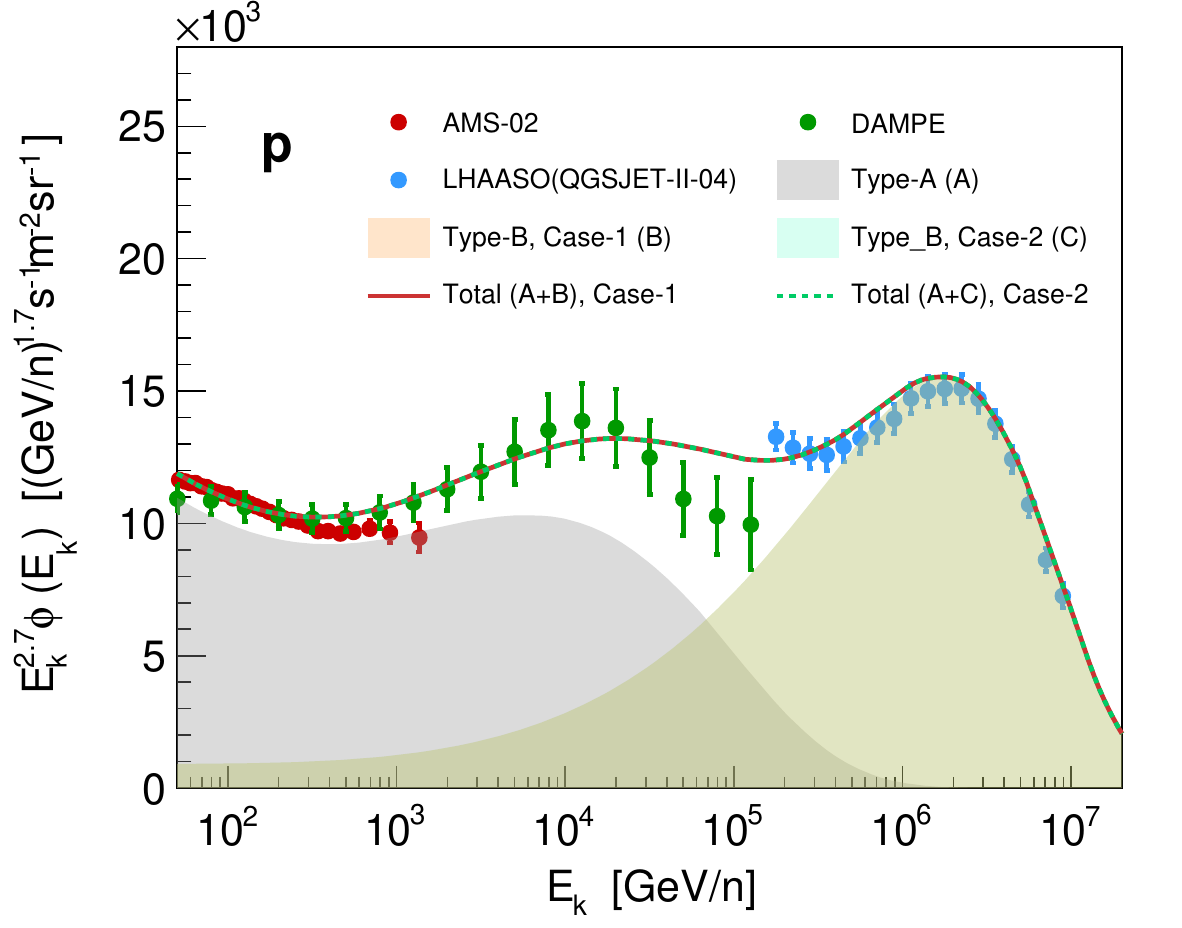}
        \label{fig:inj_ams}
    \end{subfigure}
    \hfill
    \begin{subfigure}[t]{0.46\textwidth}
        \centering
        \includegraphics[
            width=\linewidth,
            height=6.6cm,                  
            bb=0 0 500 400,              
        ]{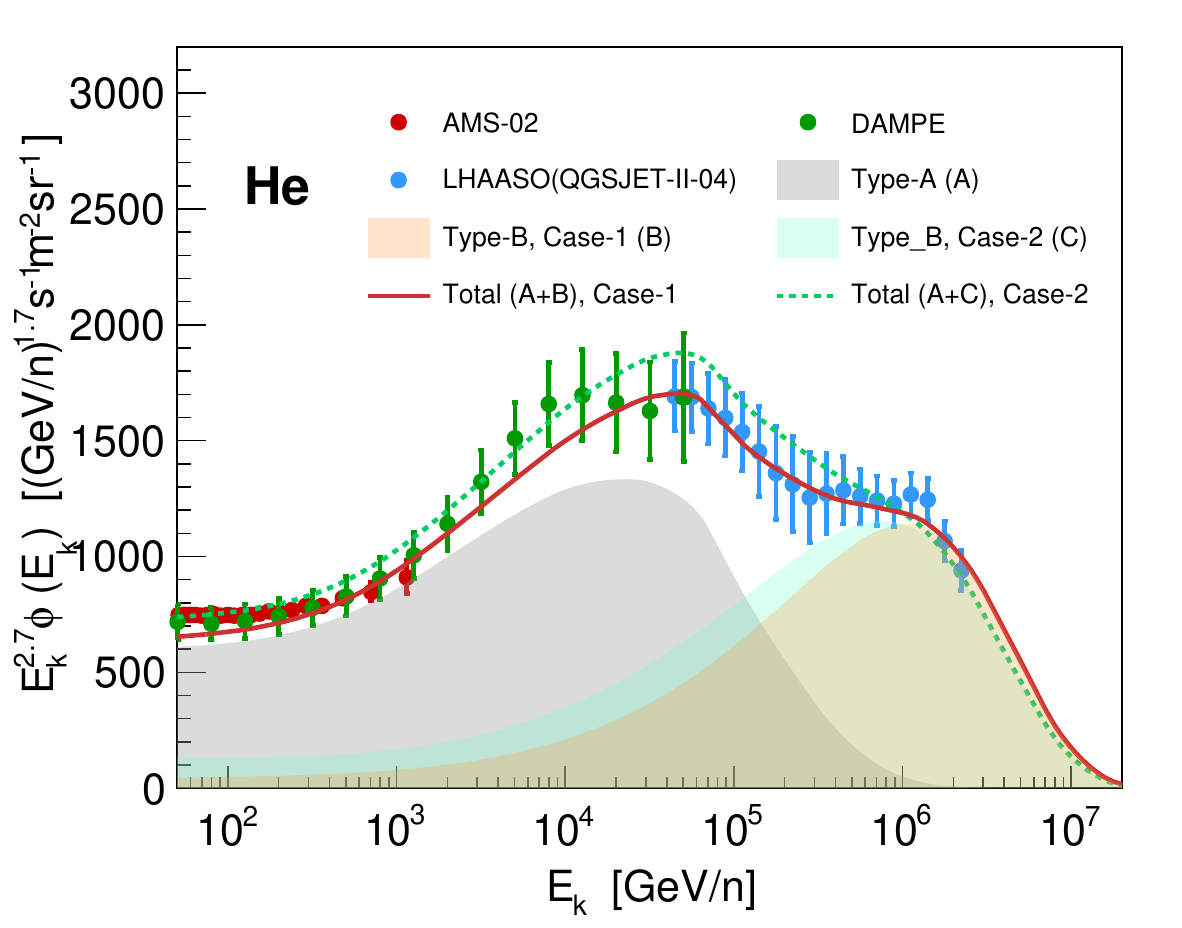}
        \label{fig:inj_dampe}
    \end{subfigure}
    \hfill

\caption{ Comparison of the model-calculated proton (left) and He (right) spectra with the AMS-02, DAMPE, and LHAASO (QGSJET-II-04) measurements. The red, green, and blue data points represent the  AMS-02\cite{aguilar2021alpha}, DAMPE\cite{2026Natur.653...52D}, and LHAASO (QGSJET-II-04)\cite{2026PhRvL.136l1001C} measurements, respectively; the gray shaded band represents the contribution of Type-A sources; and the red and cyan shaded bands represent the contributions of Type-B sources in Cases-1 and 2, respectively. The solid red and dashed green lines represent the sum of the Type-A contribution and the Type-B contribution in Cases-1 and 2, respectively.
}
    \label{fig:spectral_correlation}  
\end{figure*}
\linenumbers} 

\subsection{Data Fitting}

We adopt a stepwise fitting procedure, with the specific steps as follows:

1. B/C ratio and energy spectra below 100 TeV/n: Using the B/C ratio, proton, He, C, O, Ne, Mg, Si and Fe energy spectra from AMS-02\cite{aguilar2021alpha,aguilar2023properties,aguilar2021properties,aguilar2021properties1,aguilar2020properties} and DAMPE\cite{2026Natur.653...52D,Casilli:2025w8,DAMPE:2022jgy}, we determine the propagation parameters of the model as well as the injection spectra and cutoff positions of individual elements for Type-A sources.

2. Proton energy spectrum: Under the abundance assumptions of Case-1 and Case-2, respectively, we fit the normalization coefficients, the source fraction of both Type-A and Type-B sources, as well as the proton injection spectral index and cutoff position of Type-B sources, utilizing the proton spectrum data from AMS-02, DAMPE, and LHAASO.

3. He energy spectrum: Under each of the two abundance assumptions (Case-1 and Case-2), we adjust only the injection spectral index of helium to match the helium spectrum measured by LHAASO.

4.Given the current lack of precise measurements of individual heavy-element energy spectra (e.g., C, O, Ne, Mg, Si, Fe) by LHAASO, we adopt a reasonable approximation to reduce the number of free parameters and maintain self-consistency: we assume that the heavy elements (C, O, Ne, Mg, Si, Fe) in Type-B sources share a single effective injection spectrum (i.e., the same spectral index). Then, utilizing the all-particle spectrum measured by LHAASO, we simultaneously constrain the spectral index of this effective injection spectrum under Case-1 and Case-2, respectively. Although this approximation does not fully capture the physical reality, it provides a valid and testable benchmark model.

5.Comparison of ⟨lnA⟩: After completing the above fitting procedures, we calculate ⟨lnA⟩ under Case-1 and Case-2, respectively, and quantitatively compare them with the LHAASO measurements.

6.Impact of hadronic interaction models: repeat Steps 2–5 for the proton, helium, and all-particle spectra and ⟨lnA⟩ derived from each of the three hadronic interaction models QGSJET-II-04, EPOS-LHC, and SIBYLL 2.3d.

The relevant parameters involved in the above fitting procedures are summarized in Table 1 (propagation parameters) and Table 2 (source parameters).

{
\nolinenumbers 
\begin{figure*}[htbp]  
    \centering
    \begin{subfigure}[b]{0.46\textwidth}  
        \centering
        \includegraphics[
            width=\linewidth,
            height=6.6cm,                  
            keepaspectratio,             
            bb=0 0 500 400,              
        ]{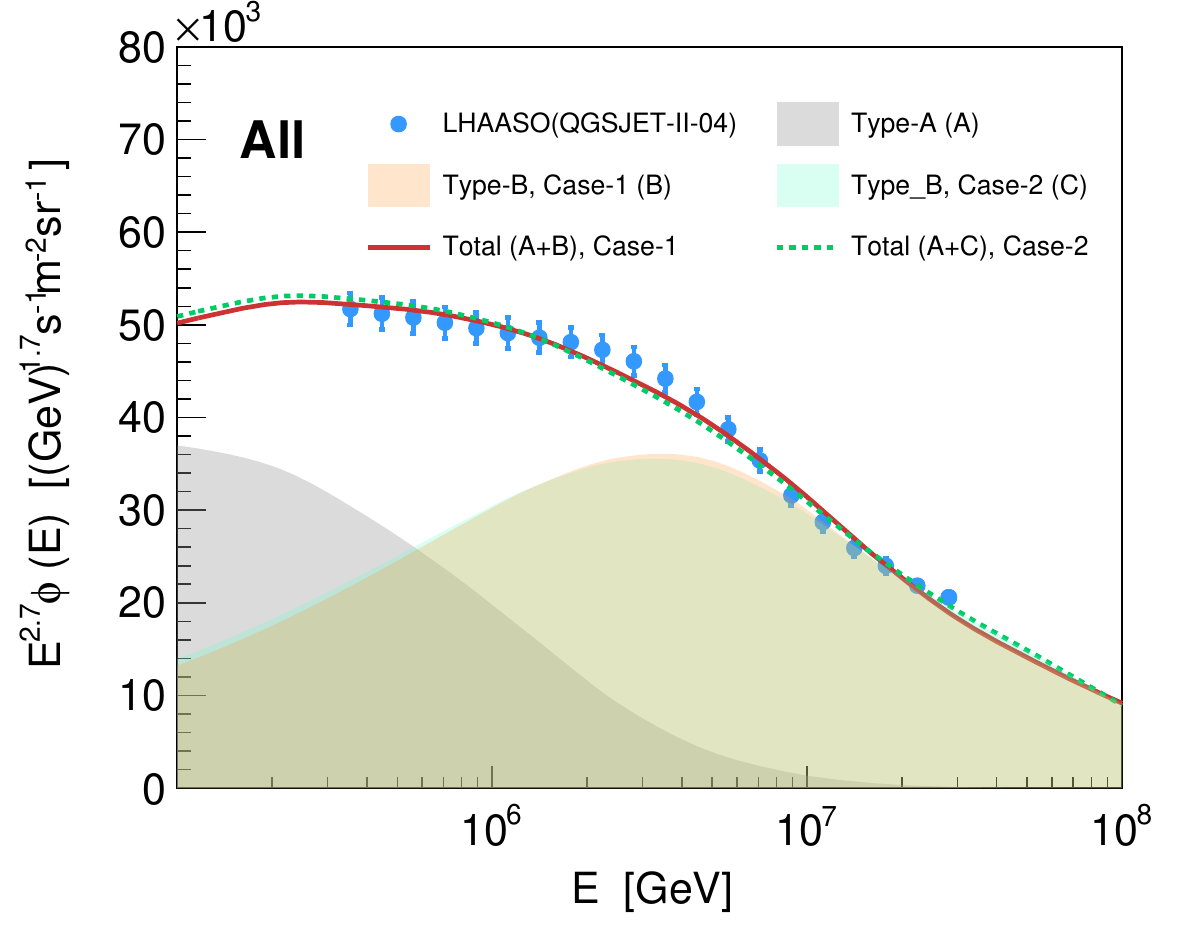}
    \end{subfigure}
    \hfill
    \begin{subfigure}[b]{0.46\textwidth}
        \centering
        \includegraphics[
            width=\linewidth,
            height=6.6cm,
            bb=0 0 500 400,              
        ]{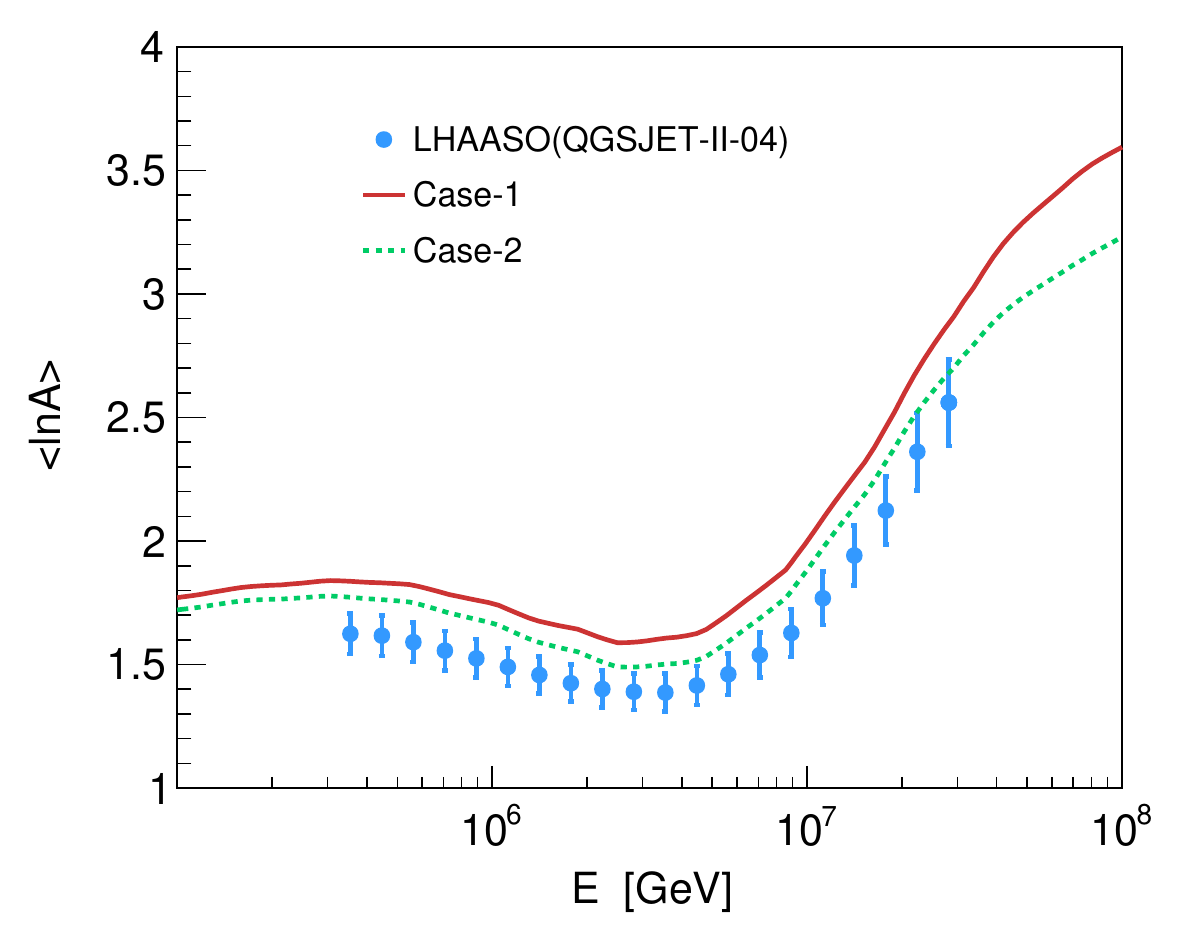}
    \end{subfigure}
\caption{Comparison of the model-calculated all-particle spectrum (left) and ⟨ln A⟩ (right) with the LHAASO (QGSJET-II-04) measurements. The blue data points represent the LHAASO(QGSJET-II-04)\cite{2024PhRvL.132m1002C} measurements. The gray shaded band represents the contribution of Type-A sources; the red and cyan shaded bands represent the contributions of Type-B sources in Cases-1 and 2, respectively. The solid red and dashed green lines represent the sum of the Type-A contribution and the Type-B contribution in Cases-1 and 2, respectively.
}
    \label{fig:spectral_correlation}  
\end{figure*}
\linenumbers} 

\section{RESULTS}

1. The Figure 1 compares the B/C ratio reproduced by our calculations based on the aforementioned method and parameters with the experimental results from AMS-02 and DAMPE. The calculated results reproduce the experimental measurements well, indicating that our choice of propagation parameters is reasonable.

2. The left panel of Figure 2 presents the comparison between the proton flux computed by the model and the experimental data from AMS-02, DAMPE, and LHAASO(QGSJET-II-04). The model calculation is based on the propagation parameters determined in Step 1, with the normalization coefficients, injection spectral indices, and cutoff positions and shapes of Type-A and Type-B sources adjusted accordingly. The model reproduces the aforementioned experimental measurements well.

3. The right panel of Figure 2 shows the calculated He energy spectra, obtained by adjusting only the helium injection spectral index under Case-1 and Case-2, respectively, based on the parameters determined in Steps 1 and 2. The results are compared with the experimental data from AMS-02, DAMPE, and LHAASO(QGSJET-II-04).

4. Building upon the aforementioned work, we constrain the effective injection spectral index of C, O, Ne, Mg, Si, and Fe under Case-1 and Case-2, respectively, utilizing the all-particle spectrum data from LHAASO(QGSJET-II-04). The left panel of Figure 3 presents the comparison between the all-particle spectrum obtained after the fitting constraints and the LHAASO(QGSJET-II-04) experimental results. The model reproduces the experimental measurements well under both abundance assumptions. 
Due to space limitations, the spectral fitting results for individual elements (C, O, Ne, Mg, Si, Fe) and their detailed comparisons with the AMS-02 and DAMPE experimental data are presented in Figure 4 of the Appendix.

5. The right panel of Figure 3 presents the comparison between the model-predicted ⟨lnA⟩ and the LHAASO(QGSJET-II-04) experimental data under Case-1 and Case-2, respectively.
Under Case-1, the model-predicted ⟨lnA⟩ is significantly higher than the LHAASO(QGSJET-II-04) measurements, exhibiting a large deviation, $\chi^2/\mathrm{d.o.f.} = 157.91/20$. In contrast, under Case-2, the model results are in good agreement with the experimental data, $\chi^2/\mathrm{d.o.f.} = 55.23/20$, $\Delta\chi^2 = 104.68$. 

6. Figures 5 presents the fitting and validation results for the LHAASO data obtained with the EPOS-LHC and SIBYLL 2.3d models, respectively. For the LHAASO (EPOS-LHC) data, under Case-1, the model-predicted ⟨lnA⟩ yields $\chi^2/\mathrm{d.o.f.} = 159.91/20$; under Case-2, $\chi^2/\mathrm{d.o.f.} = 49.71/20$, $\Delta\chi^2 = 110.20$. For the LHAASO (SIBYLL 2.3d) data, under Case 1, the model-predicted ⟨lnA⟩ yields $\chi^2/\mathrm{d.o.f.} = 80.72/20$; under Case 2, $\chi^2/\mathrm{d.o.f.} = 35.23/20$, $\Delta\chi^2 = 45.49$.

From a statistical perspective, Case-2 provides a significantly better fit to ⟨lnA⟩ than Case-1, and this conclusion is unaffected by the choice of hadronic interaction model.
This result appears to support the physical picture in which Type-B sources accelerate fully vaporized and ionized stellar material.
However, what must be noted from Tables 2 and 3 is that, within the Case-2 framework, the model is compelled to introduce large and physically unmotivated differences among the injection spectral indices of protons, helium nuclei, and other heavy-element components in order to satisfy the all-particle spectrum constraint. 
Specifically, the theory of diffusive shock acceleration (DSA) tends to predict a universal injection spectrum at a given rigidity; when various effects are taken into account, the spectral indices may differ to some degree, but this difference should be small.
When Case-2 is adopted, however, the C–Fe spectral index is harder than that of helium by 0.2—a large difference that is entirely opposite to the trend exhibited by Type-A sources (the injection spectrum becoming softer for heavier elements).

This result indicates that studies of the cosmic rays contributed by Type-B sources may currently be facing a deep degeneracy challenge between the abundance pattern and the acceleration mechanism.
This lack of physical regularity in the injection-spectrum shape suggests that the two extreme abundance assumptions and a single power-law spectral form alone may still be unable to fully reproduce the complex acceleration mechanism of Type-B sources below the knee.
This points to the need to reconsider the physical premises: whether a new abundance model should be explored, or whether the injection law of each component is itself A- or Z-dependent\citep{2026PhRvD.114d3028D}.

As for the former: given that, at present, neither new source types nor a better element-selection mechanism is available, adopting new elemental abundances is likewise devoid of theoretical and observational support. If the latter is taken, then the injection spectral index of each element itself varies with A (or Z); in that case, the contributions of “abundance differences” and “injection spectral-index differences” to the total energy spectrum cannot be distinguished, the degeneracy between the two is unavoidable, and the elemental abundances and injection spectral indices of Type-B sources are thereby difficult to determine separately.
In fact, one of our previous studies has already found, over a rigidity range from tens of GV to tens of TV, that the injection spectrum depends on A (or Z), the injection spectrum becoming softer for heavier elements; if this dependence also holds for Type-B sources, then the above degeneracy is precisely the crux of the problem.

\section{Conclusions and Predictions}
Based on the above results, we draw the following three conclusions:

1.The cosmic rays below the knee are contributed by at least two classes of sources, Type-A and Type-B: the former cuts off at a rigidity of tens of TV, and the latter at a rigidity of a few PV.

2. Using ⟨ln A⟩ as an independent test quantity, the predictions of Case-2 are significantly better than those of Case-1, and this conclusion is consistent across three hadronic-interaction models.

3. When fitting the LHAASO all-particle spectrum under the fixed Case-2 pattern, the model is compelled to introduce large and physically unmotivated differences among the injection spectral indices of the three components—protons, helium nuclei, and heavy elements (see Tables 2 and 3). Since ⟨ln A⟩ is compared as an independent test with no free parameters, these differences are a product of the all-particle spectrum fit itself, and are not introduced to accommodate ⟨ln A⟩. This indicates that the “triumph” of Case-2 cannot directly establish that its abundance pattern is correct; rather, it points to two possibilities: (i) the injection law of each component is itself dependent on A (or Z); or (ii) a new abundance model must be introduced.

On this basis, we propose a testable prediction:
If the energy spectra of individual elements can be measured precisely over the range 10 TeV/n–1 PeV/n: under Case-1, the injection spectral indices of the various elements should be close to one another; under Case-2, Mg, Si, and Fe should be significantly softer than He, C, O, and Ne. Since C–Fe share the same injection spectral index, this difference originates from the differing depth of the dip in the transition region when the Type-A and Type-B sources are superimposed, rather than from the injection spectrum itself.

The above predictions await testing through precise measurements of individual elemental energy spectra by future experiments such as HERD and LHAASO.

\begin{acknowledgments}
 

The authors thank Lin Nie, Kun Fang, and others for their substantial assistance and valuable suggestions in the completion of this work. This work was financially supported by the National Key R\&D Program of China (Grant No. 2025SKA0110103); the National Natural Science Foundation of China (Grants No. 12333006, 12275279, 12373105, and 12320101005); the Hebei Province Graduate Student Innovation Project (Grant No. CXZZBS2026083); the Open Research Fund of the Key Laboratory of Cosmic Rays (Xizang University), Ministry of Education (Grant No. 00061411/008/KLCR2026005); and the Doctoral Innovation Project of Tangshan University (Grant No. 2025065ZB).

\end{acknowledgments}
\appendix
\section{Spectra of the remaining elements, and the fit results and parameters for LHAASO (EPOS-LHC, SIBYLL 2.3d)}
{Figure 4 shows the comparison of the model-calculated energy spectra of the elements C, O, Ne, Mg, Si, and Fe with the AMS-02 and DAMPE measurements, using the parameters in Tables 1 and 2.
Figure 5 shows the results of fitting the model to the two datasets LHAASO (EPOS-LHC) and LHAASO (SIBYLL 2.3d), and Table 3 lists the parameters used in the fits.
}

{\nolinenumbers

\begin{figure*}[]  
    \centering
    \begin{subfigure}[t]{0.3\textwidth}  
        \centering
        \includegraphics[
            width=\linewidth,
            height=4.2cm,                  
            keepaspectratio,             
            bb=0 0 500 400,              
            draft=false                  
        ]{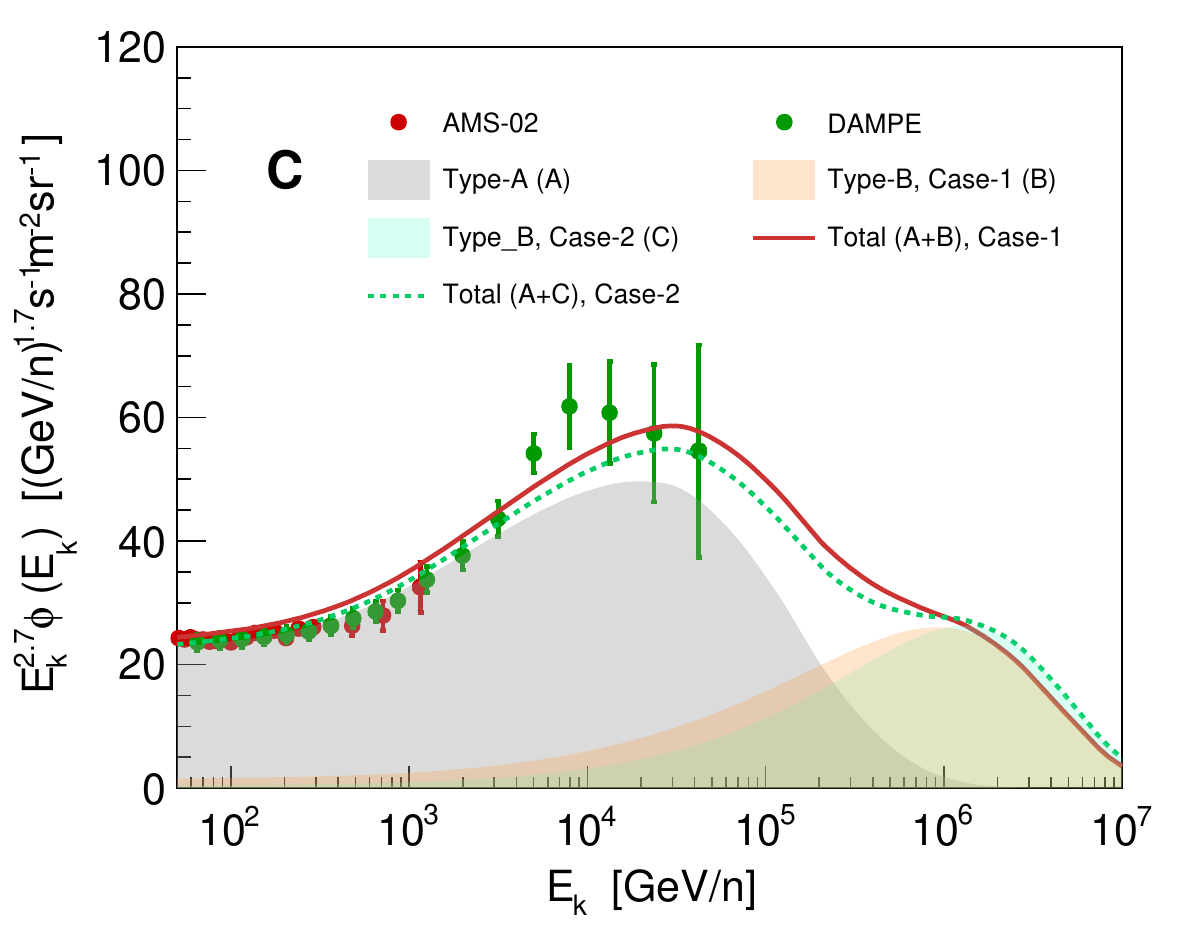}
        \label{fig:inj_ams}
    \end{subfigure}
    \hfill
    \begin{subfigure}[t]{0.3\textwidth}
        \centering
        \includegraphics[
            width=\linewidth,
            height=4.2cm,                  
            bb=0 0 500 400,              
        ]{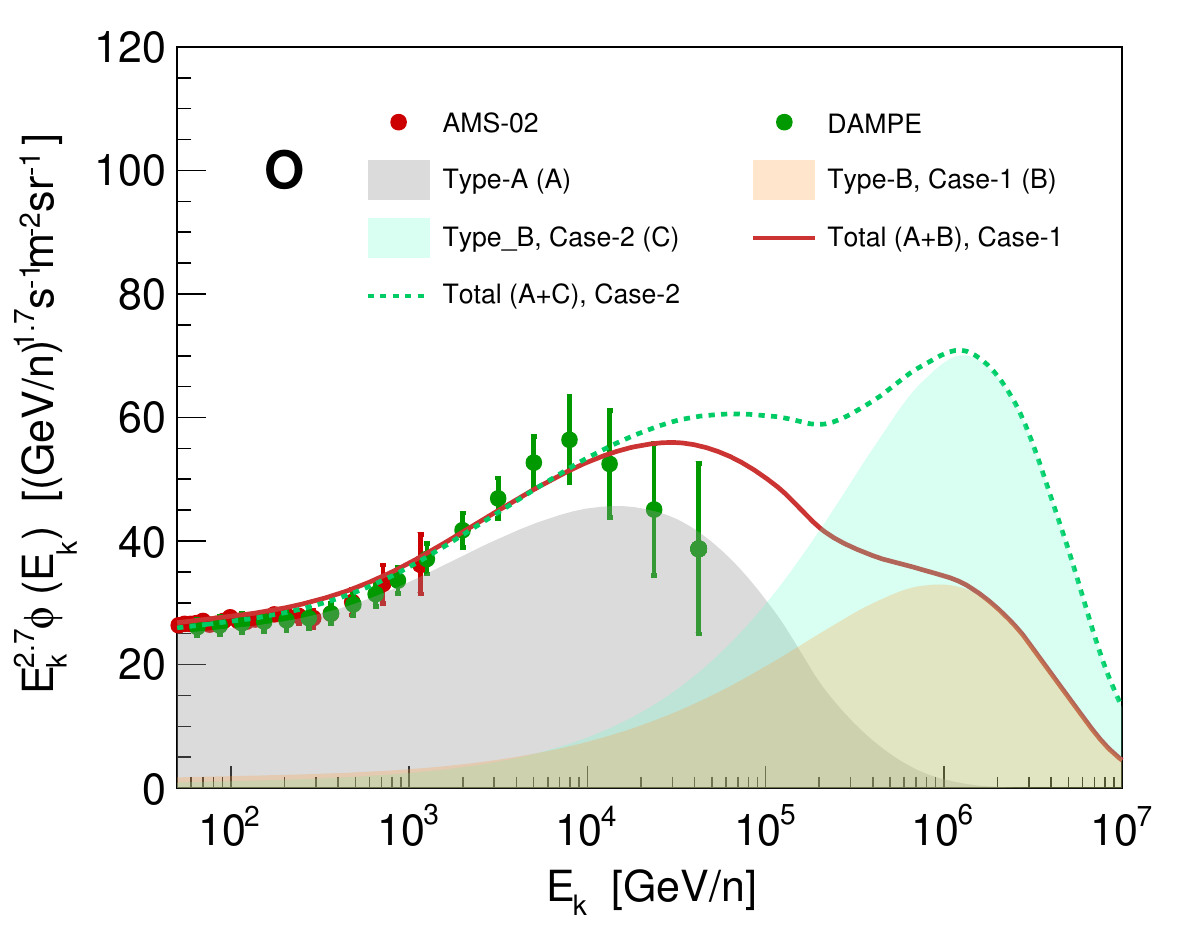}
        \label{fig:inj_dampe}
    \end{subfigure}
    \hfill
    \begin{subfigure}[t]{0.3\textwidth}
        \centering
        \includegraphics[
            width=\linewidth,
            height=4.2cm,                  
            bb=0 0 500 400,              
        ]{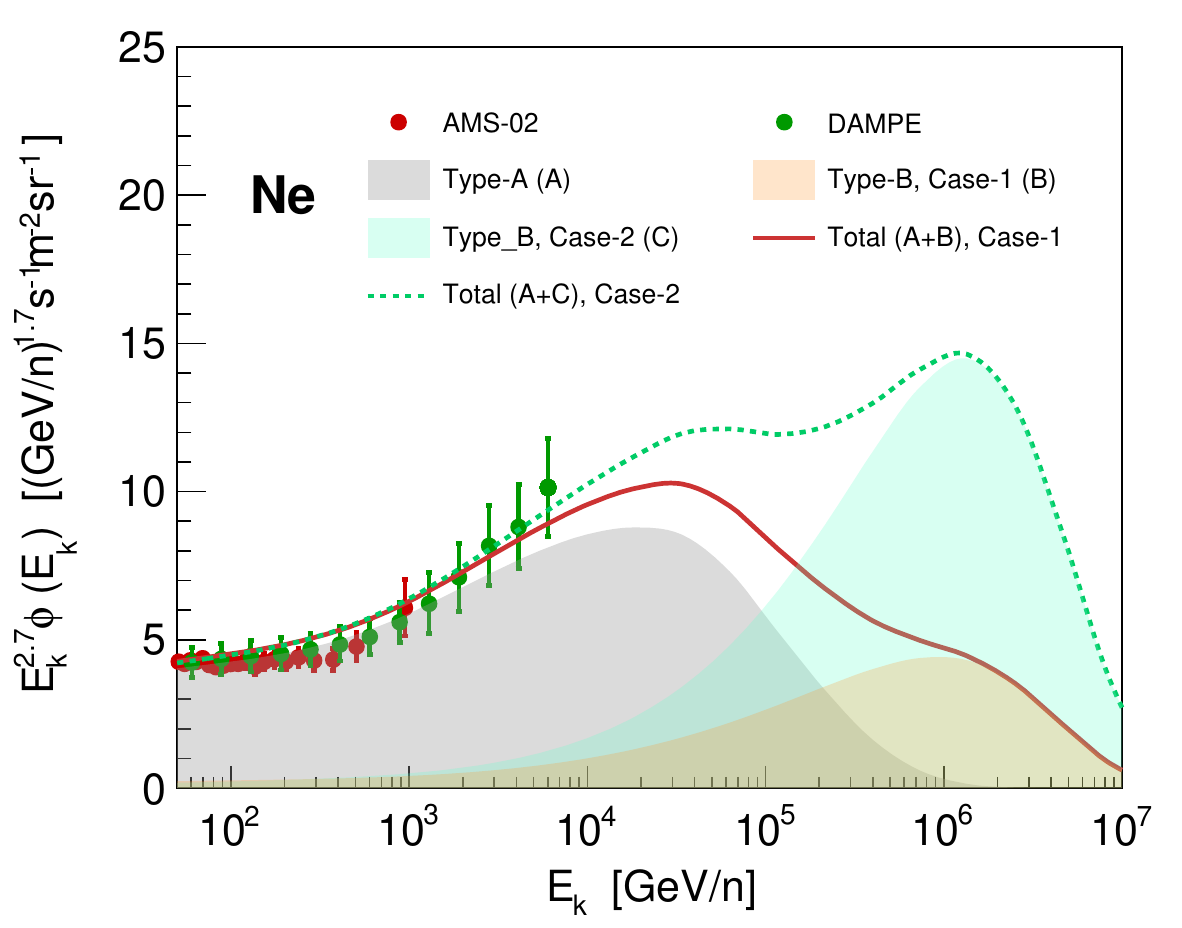}
        \label{fig:inj_dampe}
    \end{subfigure}

\begin{subfigure}[t]{0.3\textwidth}  
        \centering
        \includegraphics[
            width=\linewidth,
            height=4.2cm,                  
            keepaspectratio,             
            bb=0 0 500 400,              
            draft=false                  
        ]{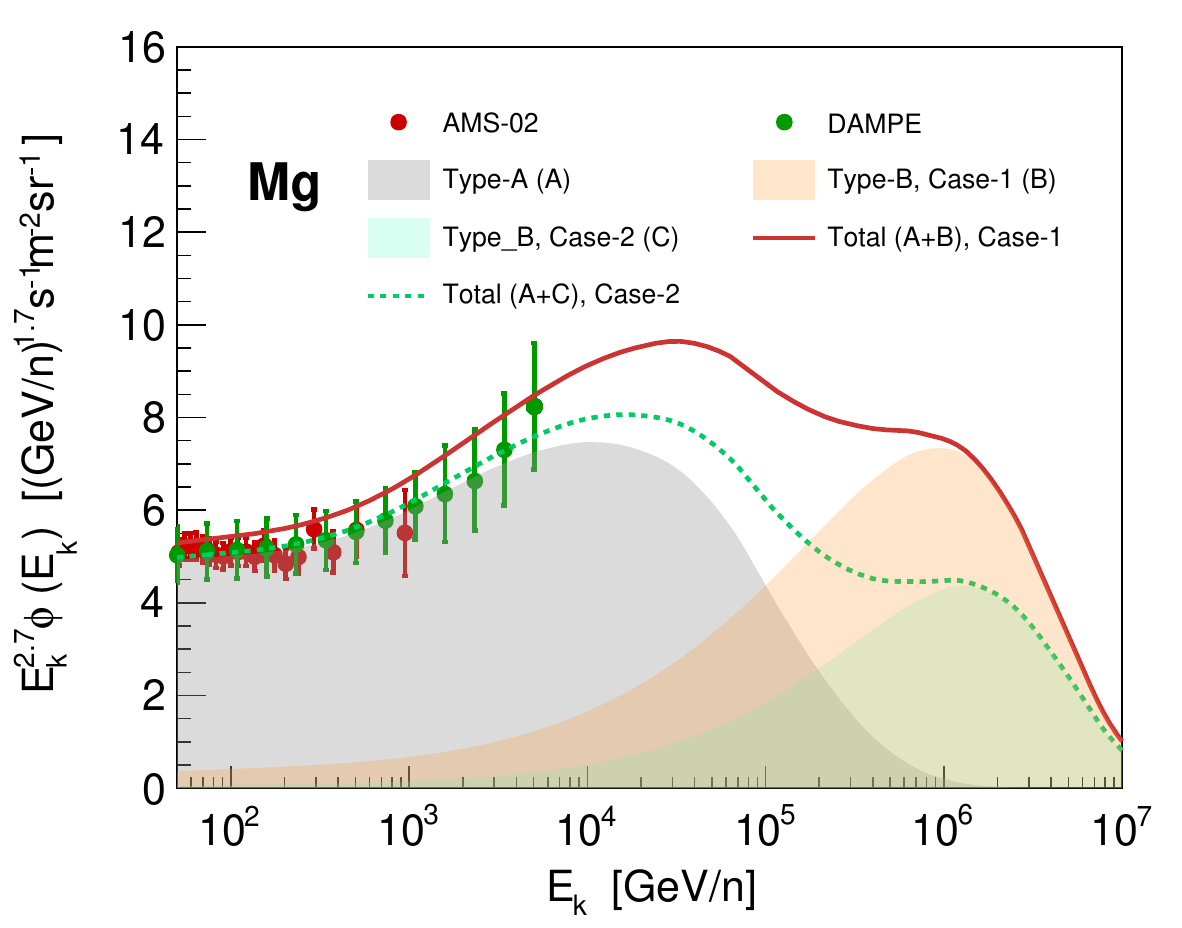}
        \label{fig:inj_ams}
    \end{subfigure}
    \hfill
    \begin{subfigure}[t]{0.3\textwidth}
        \centering
        \includegraphics[
            width=\linewidth,
            height=4.2cm,                  
            bb=0 0 500 400,              
        ]{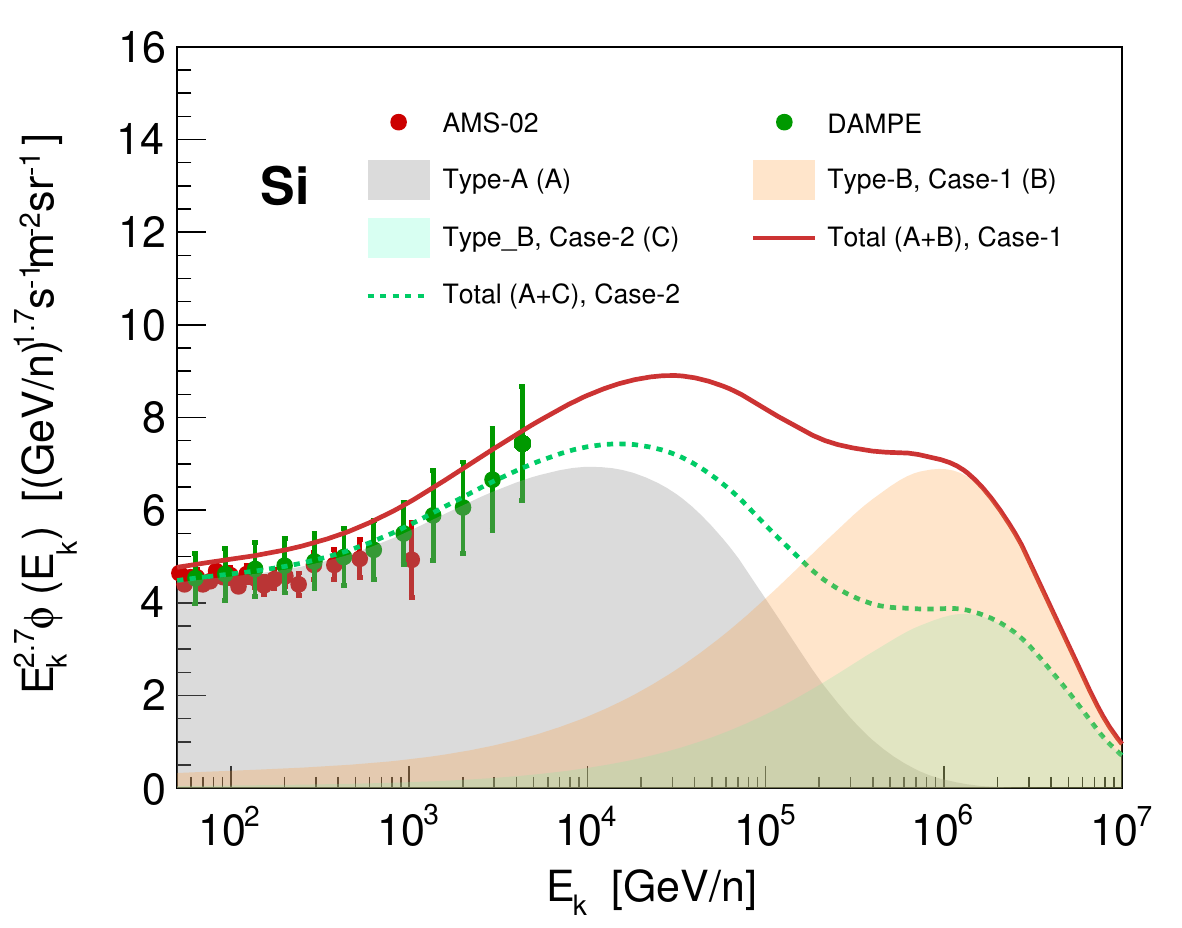}
        \label{fig:inj_dampe}
    \end{subfigure}
    \hfill
    \begin{subfigure}[t]{0.3\textwidth}
        \centering
        \includegraphics[
            width=\linewidth,
            height=4.2cm,                  
            bb=0 0 500 400,              
        ]{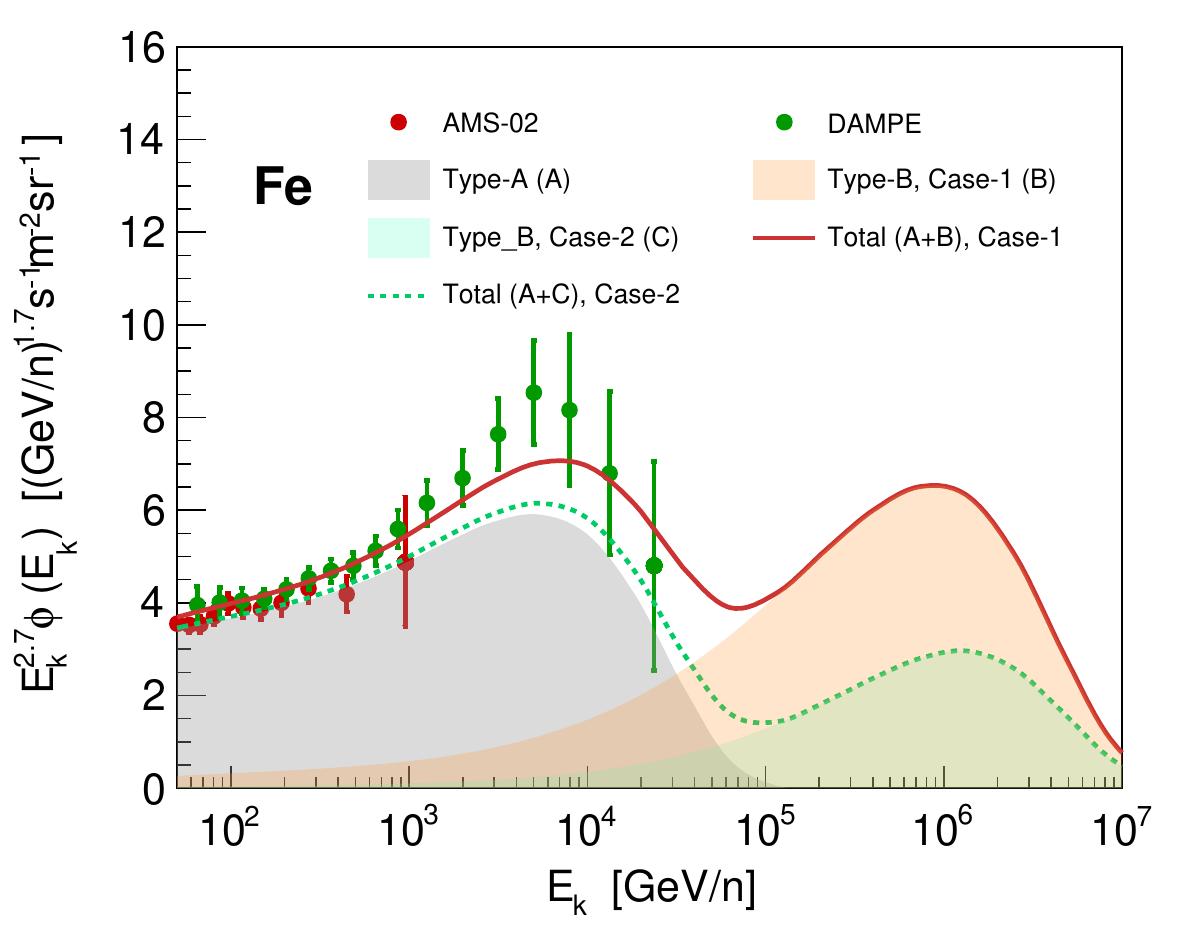}
        \label{fig:inj_dampe}
    \end{subfigure}
 
    \caption{Comparison of the model-calculated C, O, Ne, Mg, Si, and Fe spectra with the AMS-02\cite{aguilar2021alpha,aguilar2024properties,aguilar2023properties,aguilar2021properties,aguilar2021properties1,aguilar2020properties,aguilar2019properties} and DAMPE\cite{2026Natur.653...52D,Casilli:2025w8}measurements. The red and green data points represent the AMS-02 and DAMPE measurements, respectively; the gray shaded band represents the contribution of Type-A sources; and the red and cyan shaded bands represent the contributions of Type-B sources in Cases-1 and 2, respectively. The solid red and dashed green lines represent the sum of the Type-A contribution and the Type-B contribution in Cases-1 and 2, respectively.
    }
    \label{fig:injection_spectra}
\end{figure*}
\linenumbers
}

{
\nolinenumbers 

\begin{table*}[!htb]
\caption{\label{tab:table3}Spectral injection parameters.}
\begin{ruledtabular}
\begin{tabular}{lcccccccc}
& \multicolumn{4}{c}{LHAASO (EPOS-LHC)} & \multicolumn{4}{c}{LHAASO (SIBYLL 2.3d)} \\
\cline{2-5} \cline{6-9}
& \multicolumn{2}{c}{Type-B (Case-1)} & \multicolumn{2}{c}{Type-B (Case-2)} & \multicolumn{2}{c}{Type-B (Case-1)} & \multicolumn{2}{c}{Type-B (Case-2)} \\
\cline{2-3} \cline{4-5} \cline{6-7} \cline{8-9}
& Normalization & $\gamma$ & Normalization & $\gamma$ & Normalization & $\gamma$ & Normalization & $\gamma$ \\
& $[\mathrm{GeV}^{-1}\mathrm{m}^{-2}\mathrm{s}^{-1}\mathrm{sr}^{-1}]$ & & $[\mathrm{GeV}^{-1}\mathrm{m}^{-2}\mathrm{s}^{-1}\mathrm{sr}^{-1}]$ & & $[\mathrm{GeV}^{-1}\mathrm{m}^{-2}\mathrm{s}^{-1}\mathrm{sr}^{-1}]$ & & $[\mathrm{GeV}^{-1}\mathrm{m}^{-2}\mathrm{s}^{-1}\mathrm{sr}^{-1}]$ & \\
\hline
H  & $2.73\times10^{-3}$ & 2.12 & $2.73\times10^{-3}$ & 2.12 & $2.51\times10^{-3}$ & 2.12 & $2.51\times10^{-3}$ & 2.12 \\
He & $1.37\times10^{-4}$ & 2.06 & $4.97\times10^{-4}$ & 2.17 & $1.26\times10^{-4}$ & 2.04 & $4.57\times10^{-4}$ & 2.15 \\
C  & $5.71\times10^{-6}$ & 2.12 & $1.09\times10^{-6}$ & 1.98 & $5.25\times10^{-6}$ & 2.11 & $1.00\times10^{-6}$ & 1.97 \\
O  & $7.40\times10^{-6}$ & 2.12 & $2.99\times10^{-6}$ & 1.98 & $6.81\times10^{-6}$ & 2.11 & $2.75\times10^{-6}$ & 1.97 \\
Ne & $1.14\times10^{-6}$ & 2.12 & $6.26\times10^{-7}$ & 1.98 & $1.05\times10^{-6}$ & 2.11 & $5.76\times10^{-7}$ & 1.97 \\
Mg & $1.67\times10^{-6}$ & 2.12 & $1.90\times10^{-7}$ & 1.98 & $1.54\times10^{-6}$ & 2.11 & $1.75\times10^{-7}$ & 1.97 \\
Si & $1.59\times10^{-6}$ & 2.12 & $1.65\times10^{-7}$ & 1.98 & $1.46\times10^{-6}$ & 2.11 & $1.52\times10^{-7}$ & 1.97 \\
Fe & $1.52\times10^{-6}$ & 2.12 & $1.48\times10^{-7}$ & 1.98 & $1.40\times10^{-6}$ & 2.11 & $1.36\times10^{-7}$ & 1.97 \\
\end{tabular}
\end{ruledtabular}
\end{table*}

\begin{figure*}[htbp]  
    \centering
    \begin{subfigure}[b]{0.24\textwidth}  
        \centering
        \includegraphics[
            width=\linewidth,
            height=3.2cm,                  
            keepaspectratio,             
            bb=0 0 500 400,              
        ]{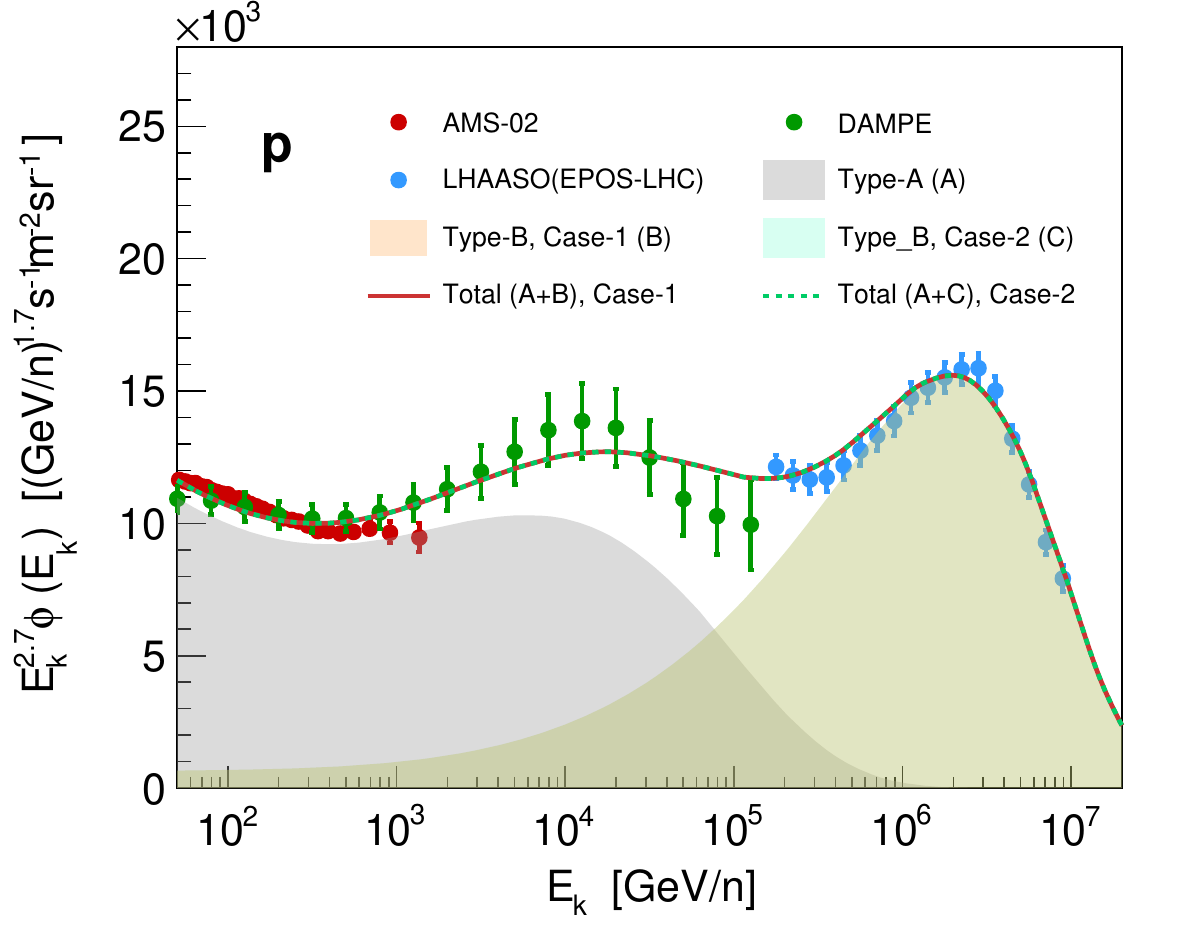}
    \end{subfigure}
    \hfill
    \begin{subfigure}[b]{0.24\textwidth}
        \centering
        \includegraphics[
            width=\linewidth,
            height=3.2cm,
             keepaspectratio,             
            bb=0 0 500 400,              
        ]{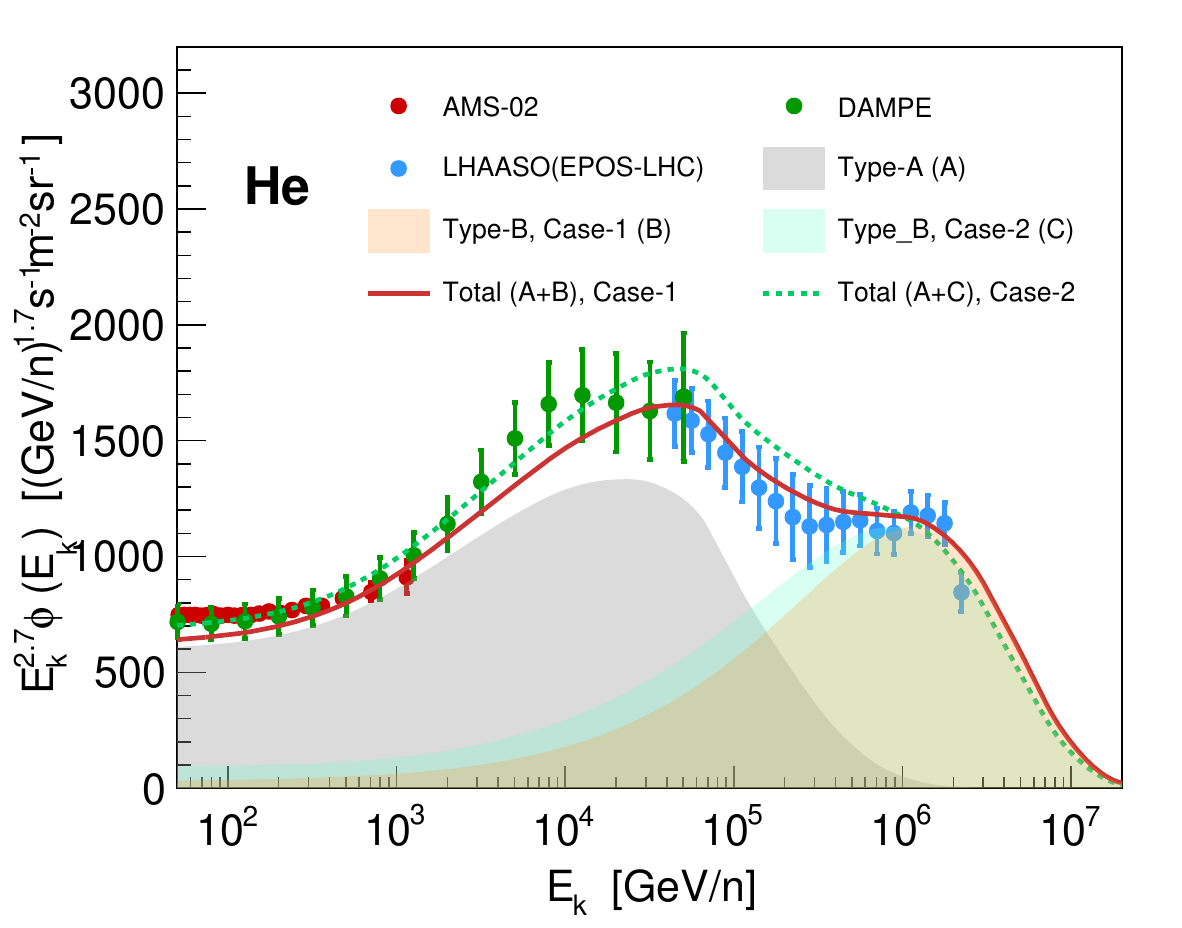}
    \end{subfigure}
    \hfill
    \begin{subfigure}[b]{0.24\textwidth}
        \centering
        \includegraphics[
            width=\linewidth,
            height=3.2cm,
             keepaspectratio,             
            bb=0 0 500 400,
        ]{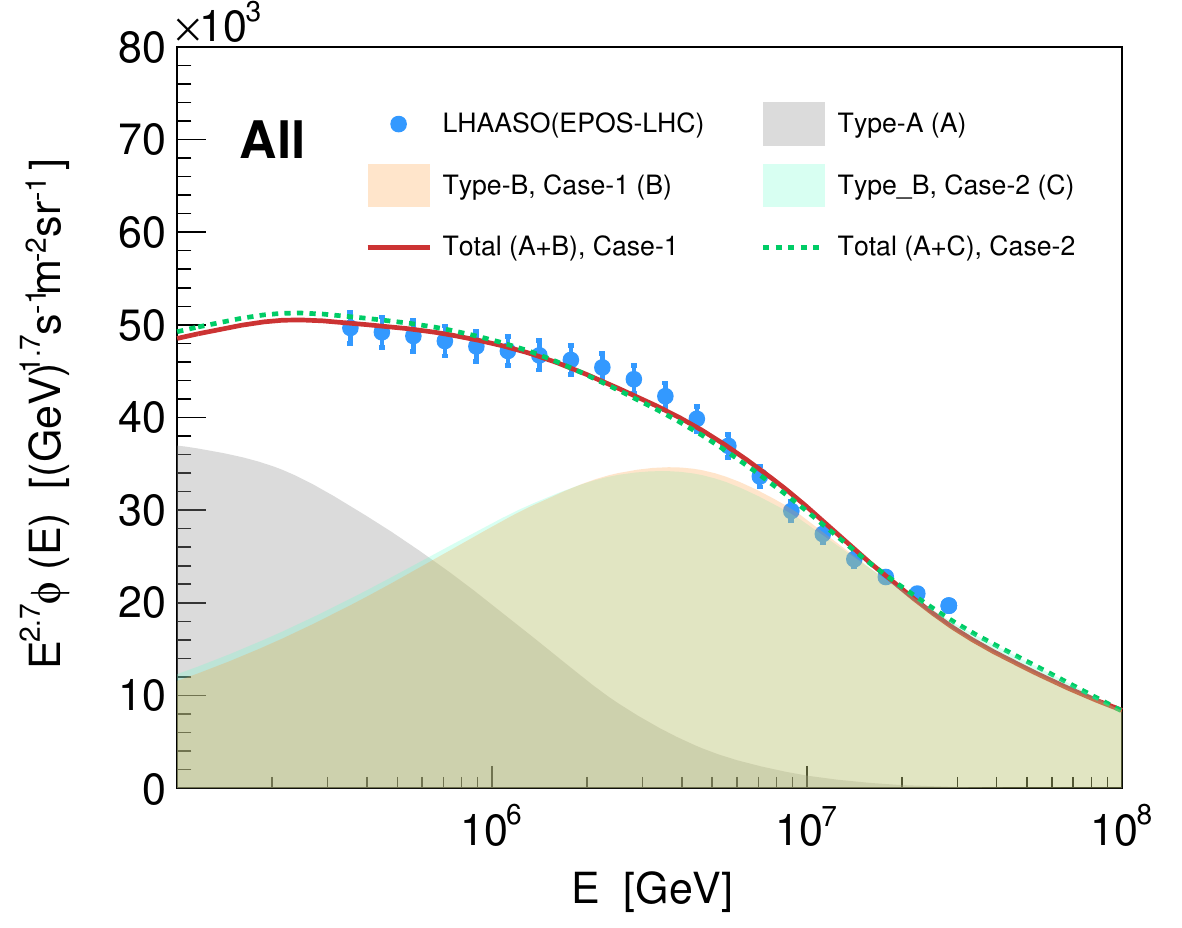}
    \end{subfigure}
    \hfill
    \begin{subfigure}[b]{0.24\textwidth}
        \centering
        \includegraphics[
            width=\linewidth,
            height=3.2cm,
             keepaspectratio,             
            bb=0 0 500 400,
        ]{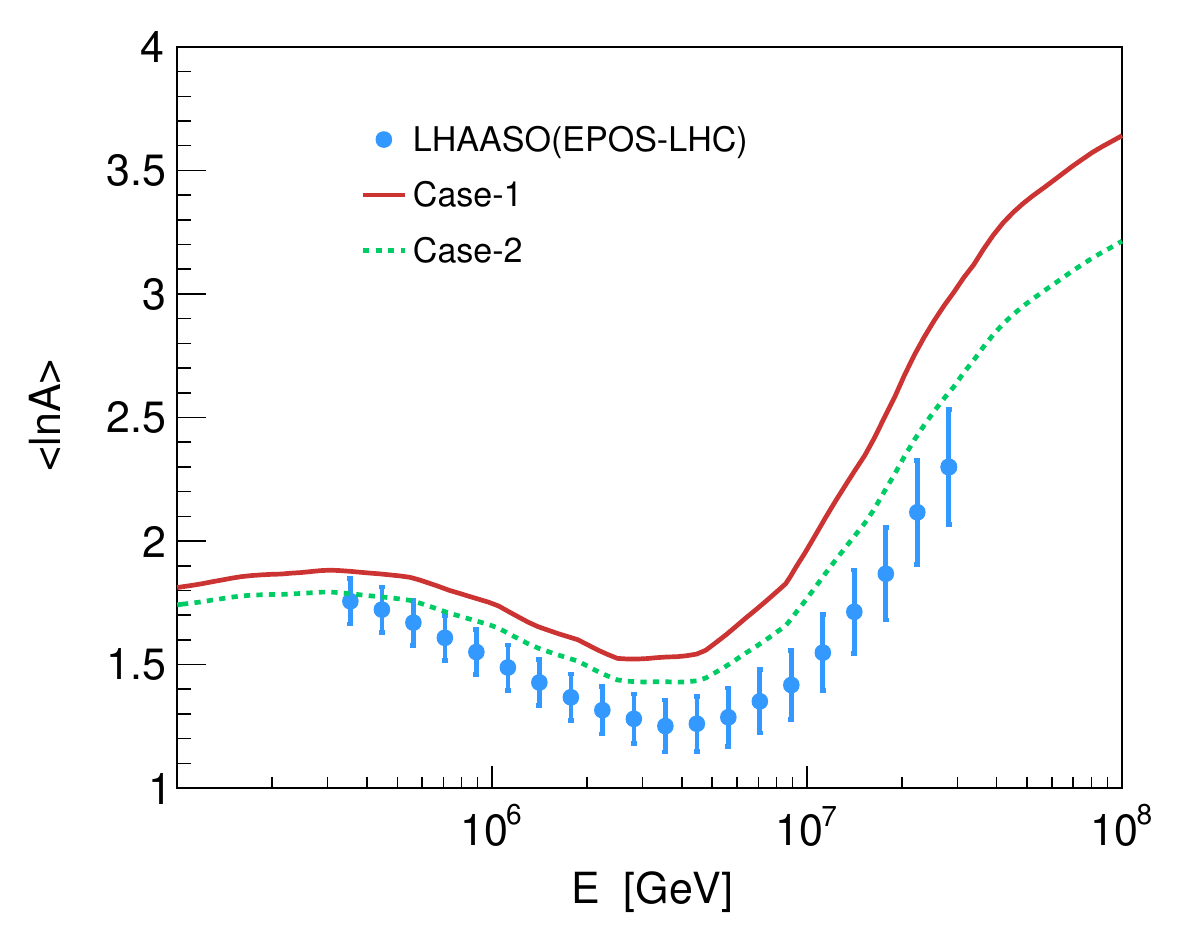}
        
    \end{subfigure} 
    
     \vspace{0.5cm}  
     
    \begin{subfigure}[b]{0.24\textwidth}  
        \centering        
        \includegraphics[            
            width=\linewidth,            
            height=3.2cm,                  
            keepaspectratio,             
            bb=0 0 500 400,              
    ]{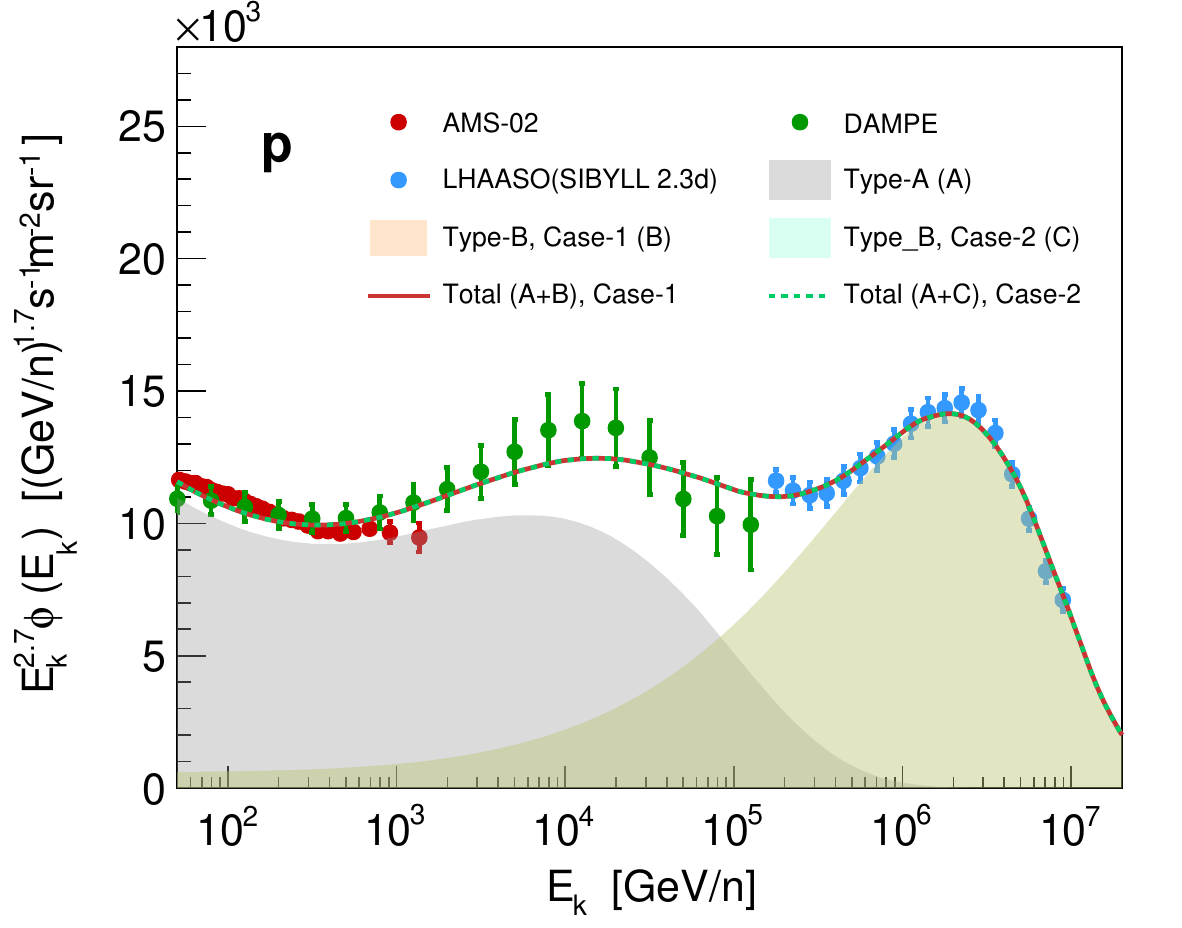}   
   \end{subfigure}    
   \hfill    
  \begin{subfigure}[b]{0.24\textwidth}        
      \centering        
      \includegraphics[            
          width=\linewidth,            
          height=3.2cm,    
           keepaspectratio,             
          bb=0 0 500 400,              
   ]{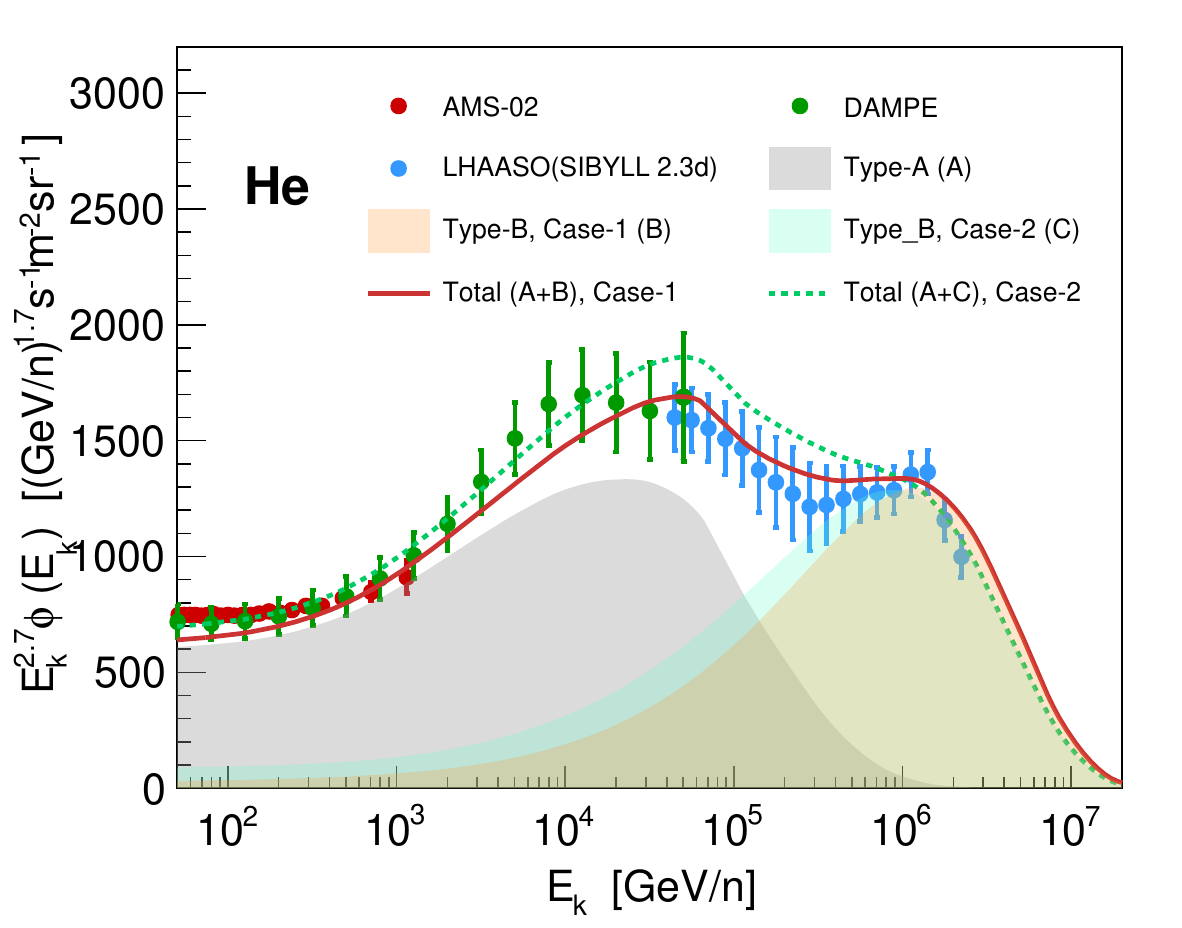}  
  \end{subfigure}            
  \begin{subfigure}[b]{0.24\textwidth}        
      \centering        
      \includegraphics[            
width=\linewidth,            
height=3.2cm,    
 keepaspectratio,             
bb=0 0 500 400,          
]{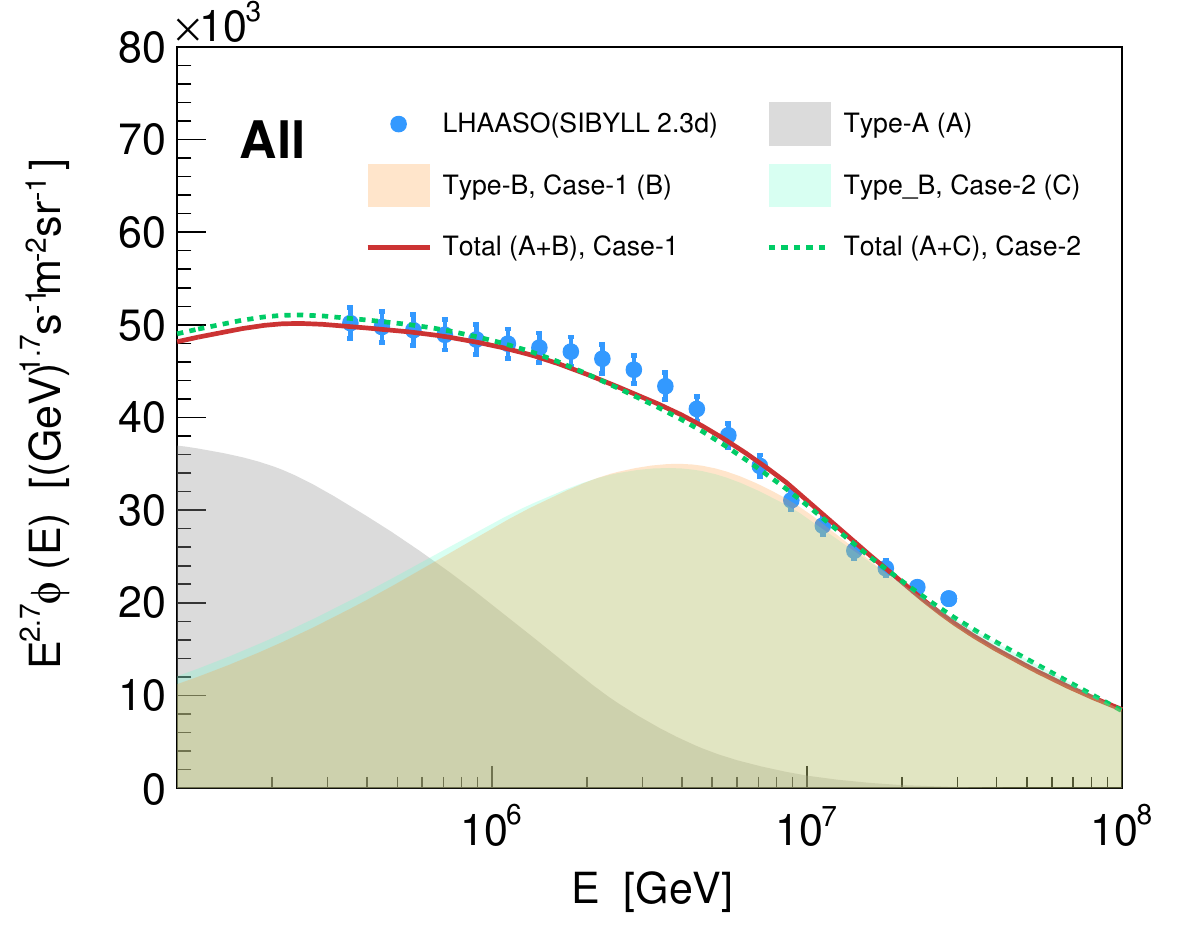} 
\end{subfigure}    
\hfill    
\begin{subfigure}[b]{0.24\textwidth}        
\centering        
\includegraphics[            
width=\linewidth,            
height=3.2cm,     
 keepaspectratio,             
bb=0 0 500 400,        
]{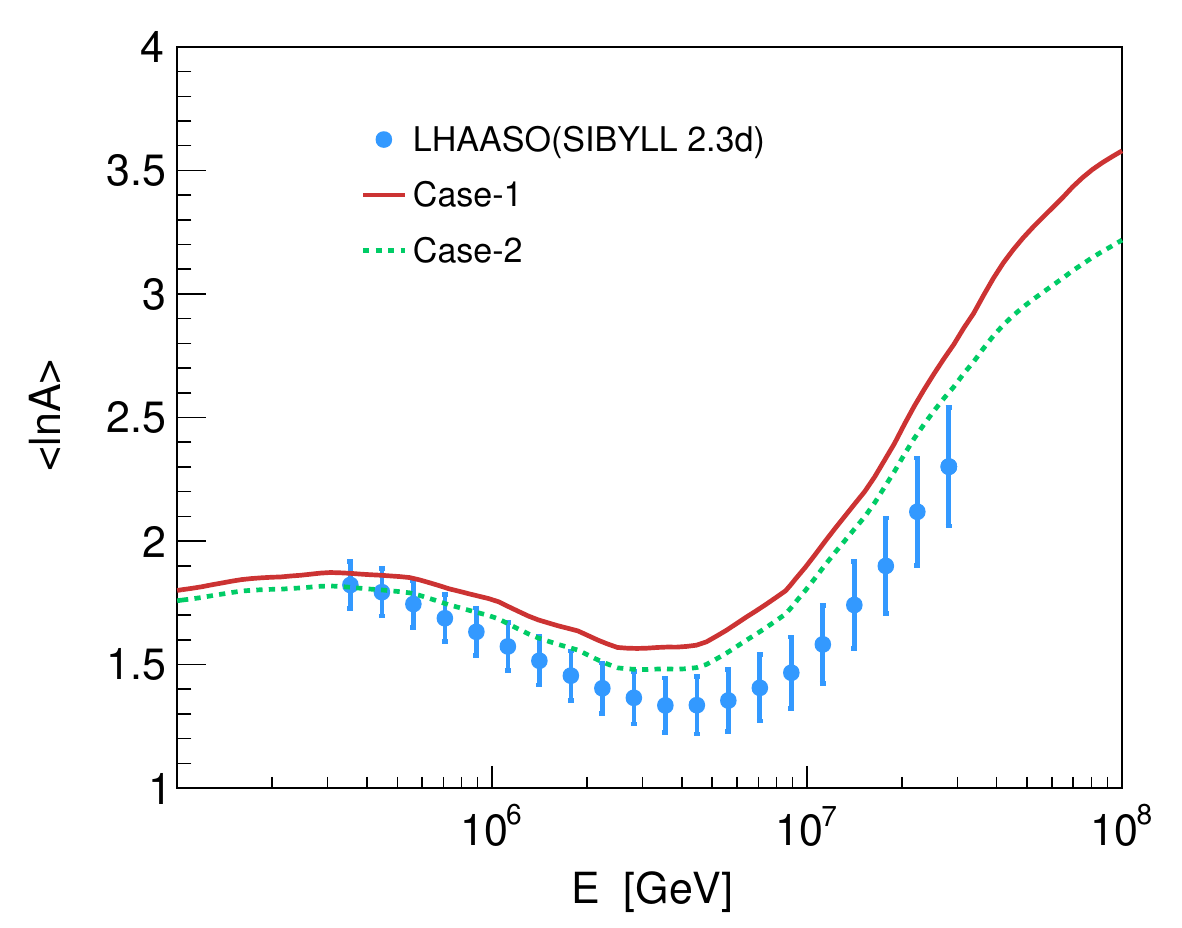}
\end{subfigure}  
\caption{
Comparison of the model-calculated proton, helium, and all-particle spectra, as well as ⟨ln A⟩, with the measurements from AMS-02, DAMPE, LHAASO (EPOS-LHC) (top row), and LHAASO (SIBYLL 2.3d) (bottom row). The red, green, and blue data points represent the AMS-02, DAMPE, and LHAASO measurements, respectively; the gray shaded band represents the contribution of Type-A sources; and the red and cyan shaded bands represent the contributions of Type-B sources in Cases 1 and 2, respectively. The solid red and dashed green lines represent the sum of the Type-A contribution and the Type-B contribution in Cases 1 and 2, respectively.
}
    \label{fig:spectral_correlation}  
\end{figure*}
\linenumbers} 

\bibliographystyle{unsrt_update}
\bibliography{apssamp}
\nolinenumbers
\end{document}